\documentclass{article}
\usepackage{epsfig}
\def\aprle{\buildrel < \over {_{\sim}}} 
\def\aprge{\buildrel > \over {_{\sim}}} 

\begin{document}     
\large

\title{CP violation  effects and high energy neutrinos}  

\author{ Paolo Lipari  \\
\small   Dipartimento di Fisica, Universit\`a di Roma ``la Sapienza",\\
\small   and I.N.F.N., Sezione di Roma, P. A. Moro 2, I-00185 Roma, Italy
~~ \\
\small   also at: Research Center for Cosmic  Neutrinos,
  ICRR, University of Tokyo}
\date{\normalsize February 5, 2001}          

\maketitle

\begin{abstract}
This  work   discusses critically the prospects of measuring
$CP$ and $T$ violation effects 
in high energy neutrino factories.
For this purpose we develop, in the standard framework with three
neutrino  flavors, simple expressions for  
the oscillation probabilities in matter that are  valid for
high $E_\nu$.
All $CP$   violating effects  vanish
$\propto  E_\nu^{-3}$ and are very difficult  to detect with high
energy neutrinos.
A significantly easier  task is the determination    of the  absolute
value $|\delta|$ of the phase in the $\nu$ mixing matrix  that controls the
$CP$ and $T$ violation effects, performing  precision measurements 
of  the $CP$ and $T$ conserving part of the oscillation probabilities.
\end{abstract}

\section{Introduction}
Measurements of atmospheric \cite{atmospheric} and  solar
\cite{solar} neutrinos have  recently given   evidence or strong
indications that neutrino oscillations  exist.
These  results, together  with important constraints from reactor
experiments \cite{reactors},  give us  precious informations  about
 the neutrino
masses and mixing that we  hope will  be of great value
to  develop an understanding on physics  beyond  the standard model.
In this work we will assume  that the oscillations are only between
the three known  $\nu$ flavors, and  the  surprising and potentially
extraordinarily important  results of LSND \cite{LSND} will be  neglected.

There is  currently a very active interest about the planning
of  future experimental studies on  $\nu$  flavor  transitions;  the
possibility  to observe   $CP$  and $T$ violation effects
in $\nu$  oscillations  (see \cite{Cabibbo,Arafune,cp-studies})
is  perhaps  the most fascinating perspective.
Neutrino  factories \cite{nu-factories}  have been proposed as a   method to
provide  intense and    well controlled  beams  
to  perform these studies.
Two fundamental  properties  of  a neutrino factory experiment
are  the energy  $E_\mu$ of the  muon beam, and the 
neutrino pathlength $L$.
Many proponents of a  neutrino  factory
experimental program  are advocating
high  $E_\mu$ \cite{nu-fact-high}, in fact as high  as technically possible
($E_\mu \sim 50$~GeV or  more)
and  long pathlength ($L \sim 3000$--7000~Km).
Other proponents  \cite{nu-fact-low,minakata} 
are advocating a  much  lower  muon energy
($E_\mu \sim 1$~GeV) and  a shorter pathlength 
($L \sim 100$~Km)  (for  the possibility
of  a  low  energy $\nu$ factory see \cite{PRISM}).
A  critical  discussion of the limits  and  merits  of the two
options is  necessary.

The study of direct  $T$--violation effects,
comparing  for example the  probabilities for the transitions
$\nu_\mu \to  \nu_e$ and $\nu_e \to \nu_\mu$, is in principle 
very  attractive \cite{tmu}, however, 
until $\nu$ beams of  extraordinary purity  become technically feasible,
this  study requires the  identification 
of the flavor and  electric  charge of  $e^\pm$; this 
is  very  difficult to do in a  very massive detector
such as those  required for  these  studies.

The  study of $CP$ violation  effects
suffers  because of   a fundamental problem:
it is  essentially 
impossible to  construct on Earth two  $CP$  antisymmetric 
long baseline experiments,
because  $\nu$'s  or $\overline{\nu}$'s  
propagate in  a medium of electrons  and  quarks
(and not positrons and anti--quarks).
The  effects of the medium   on the $\nu$  flavors transitions
are in general large, and even in the presence of
a $CP$ symmetric  fundamental lagrangian 
one will find
$P(\nu_\alpha \to \nu_\beta) \ne  P(\nu_\beta \to \nu_\alpha)$.

The   fundamental motivation of the ``low  energy option''
is to  perform the measurements
where the asymmetry induced
by the matter effects  is  neglibly small.
since the matter  effects   grow with $E_\nu$,this  requires
low energy neutrinos; however  because of the difficulty
in focusing low  energy neutrinos, the  smallness of
interaction cross sections and the difficulty of 
flavor  identifications the experimental  challenges are daunting.  

\begin{figure} [t]
\centerline{\psfig{figure=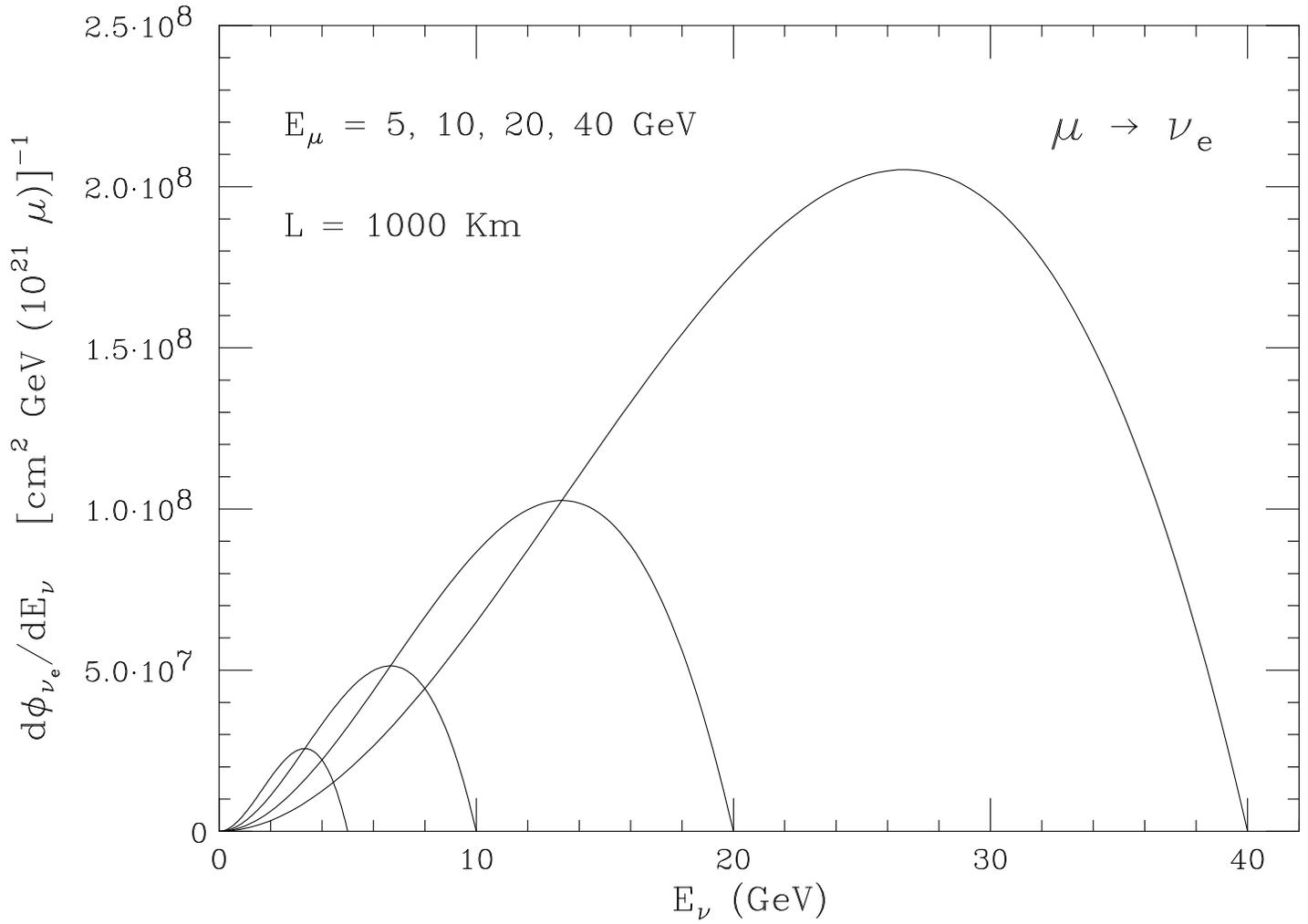,angle=90,height=13.5cm}}
 
\caption {Electron neutrino  fluence   ($d\phi_{\nu_e}/dE_\nu$) in 
a neutrino factory machine. The different  curves  are calculated for a
fixed  number of  unpolarized  $\mu$ decays, with different
energy: $E_\mu = 5$, 10, 20 and 40~GeV.
Note that   increasing $E_\mu$ the  integrated  $\nu$ fluence  
increases $\propto E_\nu^2$, but the fluence at low energy
decreases.
\label{fig:flux}  }
\end{figure}

\begin{figure} [t]
\centerline{\psfig{figure=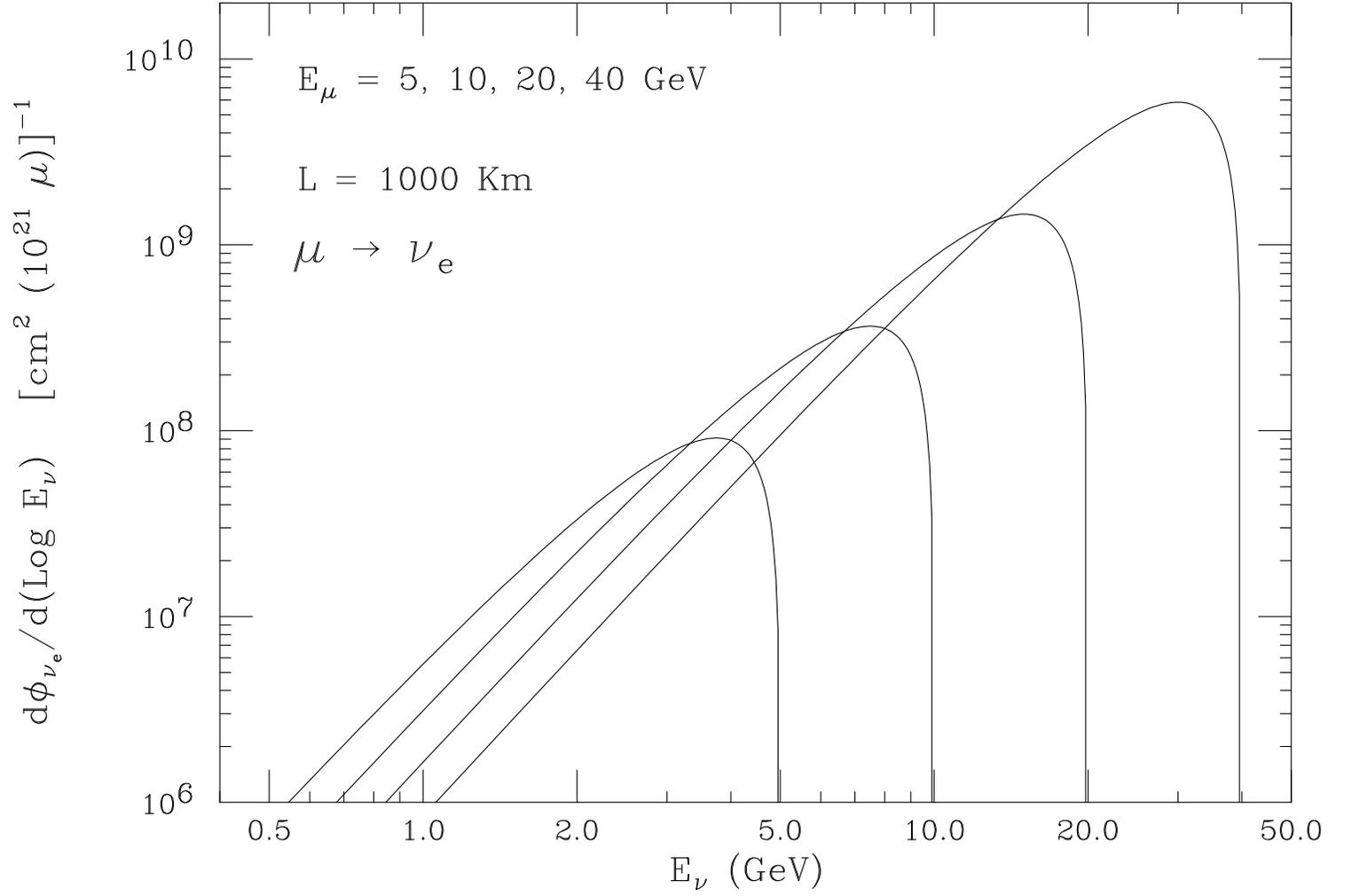,angle=90,height=13.5cm}}
 
\caption {Electron neutrino  fluence in   
a neutrino factory  machine  ($d\phi_{\nu_e}/d\log E_\nu$).
\label{fig:flux1}  }
\end{figure}

The  key point   in favour of the  choice of  a very  high
$E_\mu$   for a neutrino  factory 
is that   the   rate  of neutrino  events,  
increases   $\propto  E_\mu^3$.
This   impressively  rapid    growth of the event rate
is readily understood, as  the consequence of two  effects:
the average energy   of  the  secondary neutrinos
grows
linearly  with $E_\mu$,
  and to a good  approximation $\sigma_\nu \propto E_\nu$;
moreover the angular opening of the 
neutrino  beam  shrinks as $\gamma^{-2} = (E_\mu/m_\mu)^{-2}$,
 correspondingly the
intensity of the  $\nu$ fluence  at a far  detector increases as $ \propto
E_\mu^2$.
What is  not  often sufficiently  stressed  is that 
increasing $E_\mu$ 
one has  to pay a very high  price:
the  fluence of  lower energy neutrinos 
(for  a constant  number of  muon decays)  is suppressed 
$\propto E_\mu^{-1}$. For example    assuming  perfect  focusing, 
non polarized muon   beam, and approximating $\beta_\mu \simeq 1$
the fluence of electron (anti--)neutrinos is:
\begin{equation}
\phi_{\nu_e} (E_\nu) = { 12 \, N_\mu \over \pi \, L^2} ~
{E_\nu^2 \over m_\mu^2 \, E_\mu } \left (1 - {E_\nu \over E_\mu}
\right ) ~\Theta [E_\mu - E_\nu]
\end{equation}
where $N_\mu$ is the number of useful  muon decays,  $m_\mu$ 
is the muon mass  and $\Theta$ is the step  function  (the fluence 
vanishes for $E_\nu > E_\mu$).
Examples of the  fluence are  shown in fig.~\ref{fig:flux}
and ~\ref{fig:flux1}.  The key point is the fact that for $E_\nu$  
much  smaller  than $E_\mu$ the fluence has  the simple form
$\propto  E_\nu^2/E_\mu$.

What is  important in the experimental program
of course is  not the number of $\nu$   events
but the size of  the effects  of the oscillations on the  event  rates,
and actually still more important is the size of the {\em new}  effects
that one  want to study.
The oscillation  probabilities are suppressed  for
high $E_\nu$, and  therefore  it is  not immediately  obvious 
that  the rapid   growth of the  event  rate
with  increasing  $E_\mu$ is sufficient   to  compensate for the
 suppression of the
oscillation probability  for higher energy neutrinos.
One  should also    take into account the fact 
 that larger  $E_\nu$  means
larger  matter effects, and therefore   requires a larger
``subtraction''  to extract the fundamental $CP$  violation  effects
from the data.

The purpose  of this paper  is to 
analyse the size  of the  $CP$  violation effects  for  high energy
neutrinos. 
In this work  ``high energy''
means   the energy range where, for  a given  $\nu$  pathlength $L$,  
the transition  probabilities   decrease  monotonically  to zero
with  growing $E_\nu$. 
This  happens for
\begin{equation}
  E_\nu \aprge {|\Delta m^2_{23}| \, L
\over 2 \pi}  = 0.81~\left ({L \over 10^3{\rm km}} \right )~
\left ( {|\Delta m^2_{23}| \over  3 \times 10^{-3}~{\rm eV}^2}
\right )
~{\rm GeV}
\end{equation}
that is  for  $E_\nu$ larger than few  GeV even  for
the  longest   possible distances.
This  is  the energy  range where the proposed 
high  energy  neutrino  factories
will have most of their rate.

As we will discuss  in more  detail  later
(see eq.~\ref{eq:prob-asymptotic}),   for large $E_\nu$
the oscillation probabilities have the   following 
dominant    functional  dependences  on 
the $\nu$ energy and the pathlength: 
\begin{eqnarray}
& ~~& P_{\nu_e \to \nu_\mu}  \sim E_\nu^{-2} ~L^2 \nonumber \\
& ~~& \Delta P_{\nu_e \to \nu_\mu} (CP)  \sim E_\nu^{-3} ~L^3  \\
& ~~& \Delta P_{\nu_e \to \nu_\mu} ({\rm matter}) 
 \sim E_\nu^{-3} ~L^4 \nonumber 
\end{eqnarray}

The key point  is  that the $CP$ violation effects  vanish
rapidly with increasing  energy.  Also  important
to note is the  fact  that the
fundamental $CP$ violation   
effects  and   matter effects  have the  same   
asymptotic energy dependence,  but different dependences on 
the pathlength $L$.
Integrating these   probabilities  over the expected  energy spectrum 
for   a neutrino  factory far detector
one   finds the  following scaling  laws
for different  signals: 
\begin{eqnarray}
& ~~& {\rm Rate} \sim E_\mu^{3} ~L^{-2} \nonumber \\
& ~~& {\rm Rate}_{\nu_e \to \nu_\mu}  \sim E_\mu  ~L^0 \nonumber \\
& ~~& \Delta {\rm Rate}_{\nu_e \to \nu_\mu} (CP)  \sim  E_\mu^0 ~L 
\label {eq:rate0} \\
& ~~& \Delta {\rm Rate}_{\nu_e \to \nu_\mu} ({\rm matter})
  \sim  E_\mu^0 ~L^2 \nonumber
\end{eqnarray}
The  rate  of   ``oscillated  events''
is  approximately independent  from the pathlength $L$
and grows   linearly with $E_\mu$. 
The first effect is the result  of
 cancellation between the  decrease  $\propto  L^{-2}$  of the
neutrino  fluence, and the growth  $\propto L^2$   
of the oscillation  probabilities with increasing distance.
The $E_\mu$ dependence 
is the result  of the  combination
of the  decrease of the   oscillation probability
 $\propto E_\nu^{-2}$,
with the growth of the neutrino
fluence  and cross section.
The  contribution   of $CP$  (or $T$) violation effects on  the 
event  rate  is  however approximately independent of $E_\mu$  reflecting 
a cancellation of  of the   energy  depedence   $\Delta P_\nu(CP) \propto
E_\nu^{-3}$   with the growth of the neutrino
fluence  and cross section.
The $CP$ and  $T$ violation effects  are therefore more and more 
difficult to  observe with increasing $E_\mu$, because the
size of  the $CP$  violating effects  on the rate 
 is  constant, while  the ``background''
due to the $CP$ conserving part  of the oscillation
probability  increases  linearly  with $E_\mu$.

The simple argument that we have  outlined
is   apparently in conflict 
with the results  of  previous works
that claim \cite{cervera,arubbia} that  
the largest  the $E_\mu$ the  highest the sensitivity to
the phase $\delta$.
The reason  for this  apparent discrepancy is simple to  understand
and   quite  instructive.
The  leading term of the oscillation  probability:
\begin{equation}
P_{\rm leading} = 
P_{\nu_\mu \to \nu_e}^{(0)} =
P_{\nu_e \to \nu_\mu}^{(0)} =
P_{\overline{\nu}_\mu \to \overline{\nu}_e}^{(0)} =
P_{\overline{\nu}_e \to \overline{\nu}_\mu}^{(0)} =
A_{\rm lead}  ~ { L^2 \over E_\nu^2}
\end{equation} 
  is   equal for all  four
channels  related by a $CP$   or  $T$   operation,
however   the constant $A_{\rm lead}$ 
depends on the value of  $\cos \delta$.
The rate of oscillated  events  generate  by this  term    
is  independent  from
$L$  and  grows  linearly  with $E_\mu$, therefore in principle
the  higher the muon energy the more precisely the 
constant $A_{\rm lead}$  can be measured, and $\cos \delta$ determined.
A   significant part of the
sensitivity  (or ``reach'' in parameter  space)
for the phase $\delta$ claimed for 
high energy neutrino factories  actually
can be understood as simply as  the result
of a a very  high precision measurement  of the 
leading term in the oscillation probability.

There  are two  important  considerations  that should  however
be made.  The first one  is  that   to   transform
a measurement   of the constant $A_{\rm lead}$ into a measurement of
$\cos \delta$   requires  independent measurements  with sufficient
precision also of
 all other  parameters in the
neutrino mass matrix   (two  squared mass differences
and three angles)  and  a sufficiently precise  knowledge of
the material along the neutrino path.
A second  consideration, is that this  measurement has a fundamental
ambiguity.  All  $CP$ violating effects are proportional to
$\sin\delta$.   The leading term in the oscillation probability
is a $CP$ and $T$ invariant quantity, and in fact 
depends only on the module $|\sin \delta|$.
A  measurement of $\cos \delta$   that resulted in a  value
different  from
0 or 1,  would imply the existence  of   $CP$  and  $T$  violations    effects
in the lepton   sector, and allow  a  prediction of  their
size but not of  sign.
Of course such  a  result  would be of extraordinary importance,
however  its   limitations should  be clearly understood.

The  paper  is  organized as follows:  in the  next section
we   discuss our conventions  for the neutrino mixing  matrix,
section 3  gives a qualitative 
discussion of the ``geometrical  meaning'' of the  phase $\delta$,
section 4 discusses the $\nu$  mixing in matter,
section 5 and  6  discuss
 the $\nu$  oscillation probabilities in  vacuum and in a 
homogeneous medium.  In these  sections we develop  an expression
for the oscillation probability in matter  as a  power series in
$E_\nu^{-1}$   that   
can be  very useful for  an  understanding of the potential of
high energy machines.  Sections  7 and 8   contain a discussion and
some  conclusions.
Appendix A  contains  a detailed derivation  of the  most important
result of this  paper (eq.~\ref{eq:prob-asymptotic});
additional   material is in apppendices B and C.

\section{The neutrino mixing matrix}
We will consider in this work oscillations  among three  neutrinos.
The flavor and mass eigenstates are 
related  by a  unitary  mixing matrix $U$:
\begin{equation}
 | \nu_\alpha \rangle =   \sum_j
 U_{\alpha j} \, |\nu_j\rangle,
\end{equation}
For $\overline{\nu}$'s the mixing  is  given by 
the complex  conjugate matrix  $U^*$.
The  mixing matrix $U$  can  be parametrized in terms  of 
three  mixing  angles   ($\theta_{12}$,
$\theta_{13}$, $\theta_{23}$)  and one $CP$ violating phase $\delta$.
We will use the convention  suggested in the particle data book \cite{PDG}:
\begin{eqnarray}
U & = & \pmatrix{1&0&0\cr 0&c_{23}&s_{23}\cr 0&-s_{23}&c_{23}}
\pmatrix{c_{13}&0&s_{13} \,e^{-i\delta}\cr 
0&1&0\cr -s_{13} \, e^{i\delta}&0&c_{13}}
\pmatrix{c_{12}&s_{12}&0\cr -s_{12}&c_{12}&0\cr 0&0&1}
\nonumber \\
 & ~ & ~~~~ \nonumber  \\
 & = & \left(
\begin{tabular}{ccc}
$c_{12}c_{13}$      & $s_{12}c_{13}$   &  $s_{13}e^{-i\delta}$ \\
$-s_{12}c_{23}-c_{12}s_{13}s_{23}e^{i\delta}$ &
$c_{12}c_{23}-s_{12}s_{13}s_{23}e^{i\delta}$ & $c_{13}s_{23}$ \\
$s_{12}s_{23}-c_{12}s_{13}c_{23}e^{i\delta}$ &
$-c_{12}s_{23}-s_{12}s_{13}c_{23}e^{i\delta}$ & $c_{13}c_{23}$ 
\end{tabular}
\right)
\label{eq:mixingmatrix}
\end{eqnarray}
where we have used  the  notation $s_{jk} = \sin \theta_{jk}$ and
$c_{jk} = \cos \theta_{jk}$.
We need to  specify a  convention 
for the labeling of the mass  eigenstates. 
We  will   define the state   $|\nu_3\rangle$
as the ``most isolated''  neutrino  and 
 $|\nu_1\rangle$  as the    lightest between the
remaining   two  states.
Calling  $m_1$, $m_2$ and $m_3$ the  three  mass eigenvalues and
defining:
\begin{equation}
\Delta m^2_{jk} = m_k^2 - m_j^2
\end{equation}
we  therefore have that
$\Delta m^2_{12}$ is  positive by definition, 
 while $\Delta  m^2_{23}$  can have  both signs, moreover
$|\Delta  m^2_{23}| > \Delta m^2_{12}$.
The three  mixing  angles   are then defined in the 
entire  first  quadrant:  $\theta_{jk} \in [0, \pi/2]$, while the phase
is defined  in the  interval $\delta \in [-\pi,\pi]$. All  points 
in this  parameter space  represent
physically  distinct  solutions and    parametrize
an experimentally   distinguishable  ``neutrino world''.

For completeness we  note that  other
conventions  for the domain  of variability   of the mixing 
parameters  (for the same    parametrization of the mixing
matrix   we are using)  are possible. Since  $s_{13}$ and  $\delta$ 
enter the matrix  always in the combination
$s_{13}\,e^{-i\delta}$ and
$-s_{13}\,e^{i\delta}$, it is  possible for example to  enlarge the domain
of  definition  of $\theta_{13}$ to the  interval
$[-{\pi \over 2}, {\pi \over 2}]$, reducing the interval of
definition of  the phase:  $\delta \in [0, \pi]$;
the point  $(\theta_{13}, -|\delta|)$   of the conventions  used in this
paper  is then mapped   into the  point
$(-\theta_{13}, |\delta|)$.
It is  also  common to consider  both  signs of $\Delta m^2_{12}$ as
possible.  In this  case however   the angle 
$\theta_{12}$ varies  only in the  interval  $[0, {\pi \over 4}]$
with the point $(\theta_{12}, -|\Delta m^2_{12}|)$ of the new 
 convention mapped
into the  point (${\pi \over 2} - \theta_{12}, |\Delta m^2_{12}|$)
of our  convention. 

In the following discussion it will be 
sometimes convenient to  consider  a single 
 quantity  with the dimension of 
a  squared  mass. In these cases we will use
the  largest squared mass difference  $\Delta m^2_{23}$, as  a
dimensional 
quantity  and  the adimensional  ratio
\begin{equation}
x_{12} = { \Delta  m^2_{12} \over \Delta m_{23}^2}
\end{equation}
to obtain the  other squared mass differences:
$\Delta m_{12}^2  = x_{12} \,\Delta m_{23}^2$,
$\Delta m_{13}^2  = (1 + x_{12}) \,\Delta m_{23}^2$.
The  sign of  $x_{12}$ is  equal to the sign of $\Delta m^2_{23}$.

The Super--Kamiokande data on atmospheric  neutrinos
\cite{atmospheric} indicate that
$|\Delta m^2_{23}| $  is in the range 
2--$5 \times 10^{-3}$~eV$^{2}$  
and the  angle  $\theta_{23}$ is  close to ${\pi \over 4}$, while
$\theta_{13}$  cannot be large.
Reactor  experiments \cite{reactors}  like Chooz and Palo Verde
have obtained  stringent upper  limits on
$\sin^2 2 \theta_{13}$, that  together  with the result of SK
tell us  that $\theta_{13}$ is  small  ($\sin^2 \theta_{13} \aprle 0.05$). 
The  data on solar neutrinos  \cite{solar} can be interpreted 
as evidence for oscillations,   and give information
on the the angles  $\theta_{12}$ and  $\Delta m_{12}^2$
(with a   constraint of $\theta_{13}$, that has   to be  small
in agreement with terrestrial  experiments).
The allowed region  in the parameter  space is
  composed  of discrete  regions,  only one  of which,
the so  called  large mixing angle  (LMA) solution 
with $\Delta m^2_{12} \sim 10^{-5}$--$10^{-4}$~eV$^2$ and
$\theta_{12}$ close to  (but less than)  ${\pi \over 4}$,
  gives us  a reasonable chance to observe  
CP violation effects in   $\nu$  oscillations
in a standard three flavor  picture.

\section{The  ``geometrical''  meaning of $|\delta|$}
\label{sec:delta}
\begin{figure} [t]
\centerline{\psfig{figure=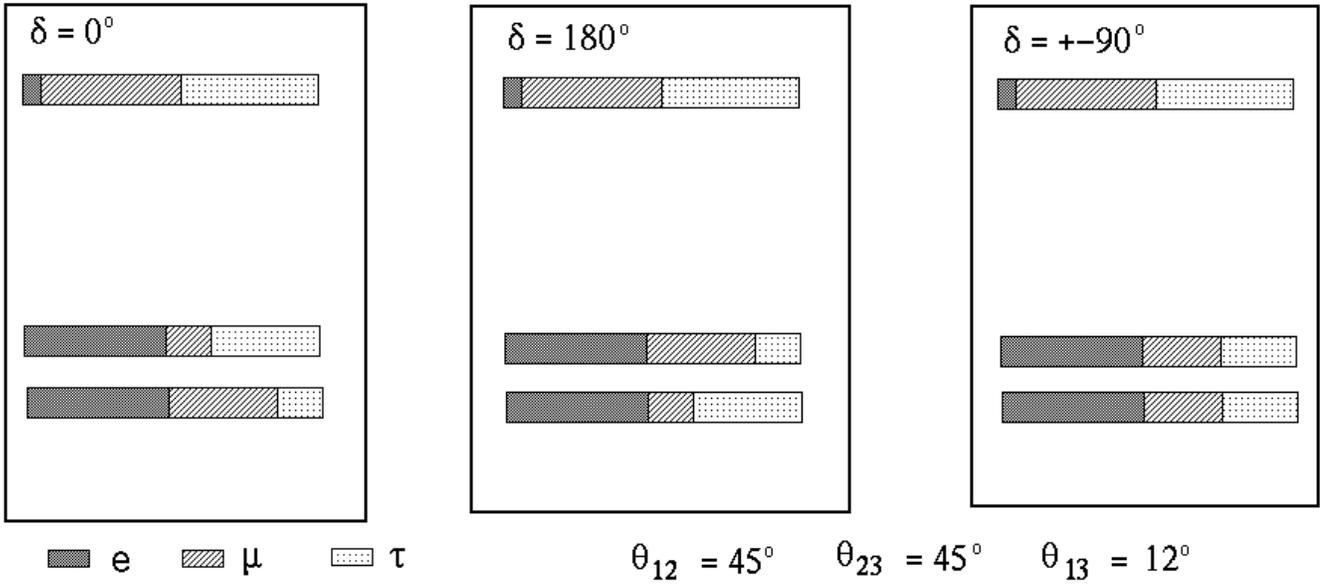,width=18.0cm}}
\caption {Graphical  representation of the  
flavor components  of the  neutrino mass eigenstates.
The  mixing angles have  the  same  values in  all three
panels: $\theta_{12} = \theta_{23} = 45^\circ$,  $\theta_{13} = 12^\circ$,
the phase $\delta$ is  0, 180$^\circ$  and $\pm 90^\circ$  in the 
left, center  and right panel.
Note how the quantities
$|\langle \nu_{\mu,\tau} | \nu_{1,2}\rangle|^2$
depend  on the value  of $|\delta|$. 
\label{fig:boxes}  }
\end{figure}

The three mixing angles 
have simple ``geometrical'' meaning in
determining  the  overlaps
$|\langle \nu_\alpha | \nu_j \rangle |^2$  
between    flavor and matter  eigenstates:
\begin{enumerate}
\item The  angle  $\theta_{13}$ 
determines how  much   electron flavor is in the state
$|\nu_3\rangle$, and how  much is  shared 
between $|\nu_1\rangle$ and $|\nu_2\rangle$
(fractions      $\sin^2 \theta_{13}$  and
$\cos^2 \theta_{13}$ respectively).
\item The   angle   $\theta_{23}$    describes  how the non--electron
content  of the state  $|\nu_3\rangle$  
is shared  between $\nu_\mu$  and $\nu_\tau$
(fractions  $\sin^2 \theta_{23}$
and   $\cos^2 \theta_{23}$).
\item The  angle  $\theta_{12}$  describes  how the   electron   flavor 
not  taken  by the $|\nu_3\rangle$ 
is shared between  the  $|\nu_1\rangle$ and the $|\nu_2\rangle$
(fractions  $\cos^2\theta_{12}$ and $\sin^2 \theta_{12}$).
\end{enumerate}
It is possible and  instructive to 
consider  also the 
``geometrical'' meaning of the absolute value of the phase
$|\delta|$  that  enters in the $CP$  conserving part of
the oscillation probabilities.
For this  purpose 
it  can be  useful  to use a  graphical  representation 
of the mixing  matrix with ``flavor boxes'',
this   representation  has  been used before
by several authors in particular by A.~Smirnov \cite{Smirnov-nu2000}.
Some examples of this  representation 
are shown in  fig.~\ref{fig:boxes}.
Each panel in   the figure  shows 
the   flavor  content of the three neutrino mass eigenstates.
In the  three panels
the  values of the three mixing  angles is  identical,
 but  the value 
of the phase $\delta$ changes.
The phase $|\delta|$ determines  how 
the muon  and tau  flavor not taken  by the  $|\nu_3\rangle$ is shared 
between the $|\nu_1\rangle$ and the $|\nu_2\rangle$.
For $\delta = 0$  the  $|\nu_1\rangle$   has the largest 
(smallest) $\nu_\mu$  ($\nu_\tau$) 
component, 
while for  $\delta = \pi$ the situation is  reversed.
It is   easy to see that the $|\langle \nu_\mu |\nu_1\rangle|^2$
overlap  grows  monotonically  with $|\delta|$, while 
the $|\langle \nu_\tau |\nu_1\rangle|^2$
overlap  decreases   monotonically, 
and the opposite  happens for $|\langle \nu_{\mu,\tau}|\nu_2\rangle|^2$.
The  range of  variation  is determined  by the values  of 
the three angles
$\theta_{12}$, $\theta_{23}$  and $\theta_{13}$.

In conclusion: the  set of overlaps  $|\langle \nu_\alpha|\nu_j\rangle|^2$
(that  is a complete solution  for the flavor  boxes) 
is equivalent to  a  perfect determination of the three
mixing  angles  {\em and} of  the value of  the phase $\delta$,
but with an ambiguity  of sign.
The mass--flavor   overlaps 
can  be determined   without  ever   measuring any $CP$  or $T$ 
violation effects, and therefore the absolute value 
 $|\delta|$  can be measured   without  observing any  such effect.
Of course mathematical   consistency imply that, 
if the determination of $|\delta|$  differs  from the 
special values  0 or $\pm \pi$, then $CP$ and $T$ violation
effects  must exist,  and we can predict   their  existence and their
 {\em size} but not their {\em sign}.

\subsection{Quasi--bimaximal mixing}
To illustrate  our  discussion with a concrete example 
it can be instructive  to consider in more  detail the cases
of  ``bimaximal''  and ``quasi--bimaximal'' mixing.
Bimaximal mixing   corresponds  to the values:
$\theta_{12} = \theta_{23} = {\pi \over 4}$ and  $\theta_{13} = 0$,
for the mixing  angles. The mixing matrix takes then the form:
\begin{equation}
U=\left(
\begin{array}{ccc}
 {1\over \sqrt{2}}       &  {1\over \sqrt{2}}      &   0  \\
- {1 \over 2}  &   {1 \over 2}     &   {1\over \sqrt{2}}   \\
+ {1 \over 2}  & - {1 \over 2}  &     {1\over \sqrt{2}}  
\end{array}
\right)
\end{equation}
It is well known that in this  case  the phase $\delta$ is
physically irrelevant.
In quasi--bimaximal mixing
we allow  for a small non vanishing  value of $\theta_{13}$.
In first order, that  is  neglecting
$\theta_{13}^2$  and approximating $c_{13} \simeq 1$,
the mixing matrix becomes:
\begin{equation}
U=\left(
\begin{array}{ccc}
 {1\over \sqrt{2}}       &  {1\over \sqrt{2}}      & 
  s_{13}\, e^{-i\delta}  \\
- {1 \over 2} (1+ s_{13}\, e^{i\delta})   &   {1 \over 2}  (1- s_{13}\,
 e^{i\delta})
    &   {1\over \sqrt{2}}   \\
+ {1 \over 2}  (1- s_{13}\, e^{i\delta})
 & - {1 \over 2}  (1+ s_{13}\, e^{i\delta}) &     {1\over \sqrt{2}}  
\end{array}
\right)
\end{equation}
Note that the   values of
 $|U_{\mu 1}|$,
 $|U_{\mu 2}|$,
 $|U_{\tau 1}|$ and
 $|U_{\tau 2}|$ are  not  fully determined  by the mixing angles
but can  vary in the interval 
$[(1 - s_{13})/2,(1 + s_{13})/2]$.  
The  extreme values in the interval are reached  when 
$\delta = 0$ or $\pm \pi$ and the matrix  is  real.
It is interesting  to observe  that  ``maximum symmetry'' (when the four
 elements
$|\langle \nu_{\mu,\tau}|\nu_{1,2}\rangle|$ are all equal)  is obtained
when $\delta = \pm {\pi \over 2}$  and  $CP$ and $T$ violating effects
 are largest.

\begin{figure} [t]
\centerline{\psfig{figure=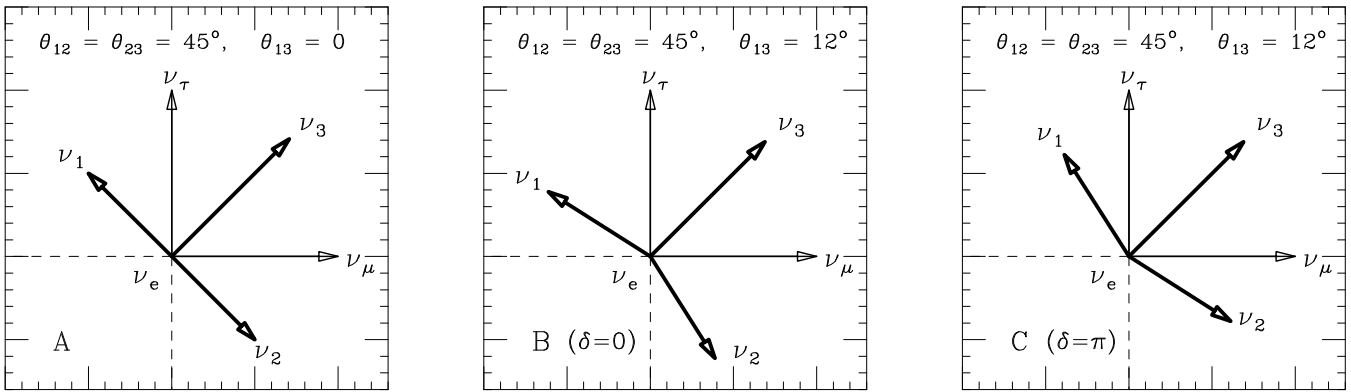,angle=90,width=18.0cm}}
\caption {Geometrical relation between the
flavor and mass eigenvectors in the case 
of a real mixing  matrix  ($\delta =0$ or $\pi$).
The  figure  shows  the projections     in the 
($\nu_\mu$,$\nu_\tau$)  plane  of the   mass eigenvectors.
The  $\nu_e$ vector  is coming out of the plane of the  figure.
The  parameters   of the mixing matrix  are indicated in the plots.
See text  for more discussion.
\label{fig:triad}  }
\end{figure}

For  an understanding of the difference  between the cases
$\delta = 0$ and
$\delta = \pi$  it can be instructive to look
at fig.~\ref{fig:triad}. The  figure 
describes 
bimaximal and  quasi--bimaximal
mixing  when    $U$  is  a  real   orthogonal  matrix.
In this case 
the flavor  and  mass eigenstates states  can be   
represented as two
sets of orthonormal vectors
in ordinary 3D space.
The three panels 
in fig~\ref{fig:triad}
show the projections in the ($\nu_\mu$,$\nu_\tau$) plane of the mass
eigenvectors.
The  $\nu_e$ axis  
is   orthogonal   to the plane of the paper
coming out  toward the reader.
The left panel
represents the case of bimaximal  mixing:
the $\nu_3$  lies  at 45$^\circ$ in the  $(\nu_\mu, \nu_\tau$) plane
while  the vectors  representing  $\nu_1$ and $\nu_2$
are 
at  45$^\circ$  with respect to the    $\nu_e$ axis.
coming out of the plane  of the figure.
The center and right panels  show  the    projections of the vectors
when  $\theta_{13}$  is  different from zero.
In the center panel  $\nu_3$  has  a small component  parallel 
to $\nu_e$, that is ``out'' of the plane of the figure
(this  corresponds  to $e^{i\delta} = 1$ or  $\delta = 0$); in the right
panel 
$\nu_3$ has  as  small component  opposite to the $\nu_e$  direction,
(``into'' the plane, corresponding  to  $e^{i\delta} = -1$ or $\delta = \pi$).
Performing a small rotation
of the mass eigenvectors  from the  situation  of the left panel,
one can easily see  the effects  on the flavor components of the
other mass   states.
For $\delta = 0$  (middle panel)
the  $\nu_1$ has (in absolute value) an overlap with
$\nu_\mu$ ($\nu_\tau$)  larger  (smaller) than   ${1\over 2}$, 
and  viceversa for  $\nu_2$.
When $\delta = \pi$, (right panel) the reverse  happens.
Allowing the matrix   $U$ to be complex, these two discrete  solutions
become  the two  extreme cases  of a continuum of  possible
different mixings.

\subsection{Boxes and  Triangles}
A  graphical  description of the 
available information on the neutrinos mixing  
has  been introduced  by G.~Fogli  and collaborators
in the form of ``triangle plots'' \cite{fogli-triangles}.
A ``solar  triangle''  plot  describes the information 
about the mass components  of $|\nu_e\rangle$:
\{$|U_{e 1}|^2$,
$|U_{e 2}|^2$,
$|U_{e  3}|^2$\}, while an
``atmospheric  triangle''  plot  describes the  information
about the  flavor components  of
  $|\nu_3\rangle$: \{$|U_{e 3}|^2$, $|U_{\mu 3}|^2$, $|U_{\tau  3}|^2$\}.
Each  plot  represents a mapping between 
the  values of the   three   $\nu$ components  and the points inside
an  equilateral  triangle, 
each   component  being proportional to the distance from the
point to  a side of the triangle.
The unitarity   constraints:
\begin{equation}
\sum_\alpha |U_{\alpha j}|^2 = 1
,~~~~{\rm and} ~~~~
\sum_j |U_{\alpha j}|^2 = 1
\end{equation}
are automatically   satisfied  since  from elementary  geometry we know
that the sum  of the  three distances  is a constant. 
The  triangle plots allow  to   indicate graphically the allowed region 
for the three components. The element $|U_{e 3}|^2$ is present
in both plots  and therefore there is a consistency check  between
the  analysis of solar  and   atmospheric  experiments.
The allowed  regions in the solar
solar and atmospheric  triangles  carry no 
in information about $\delta$, in the sense that 
when the allowed regions   in both  plots 
shrink to a single point,  this   is  equivalent to an
infinitely  precise   determination   of the three mixing angles,
with  {\em no} information about $\delta$ \cite{triangles}. 

More in  general one  can define
six different   ``triangle'' plots,  corresponding to the 
three  rows  and columns   of the mixing matrix,
that is to the mass  (flavor)  components of each flavor (mass)
eigenstate.
The set of any choice of {\em  three}  (or  more)
triangle plots (with one case equivalent to
the set of three ``flavor boxes'')
is sufficient to describe  (with  redundancy) 
all  four mixing parameters including $|\delta|$, 
leaving however  ambiguous the sign  of the phase.
To  account for the  sign  of $\delta$ 
(that is   measurable only  with the
direct  observation of $CP$  violation effects), one of the triangles
must be ``doubled''.

\section{Neutrino masses and mixing in matter}
The propagation of  neutrinos  in  a medium, 
differs from the vacuum  case. The effects
of the medium can  be taken into account 
\cite{MSW} considering an effective
potential  that is  independent from the $\nu$ energy.
In the study of flavor transitions  only the  difference between
the potentials  for  different  flavors is  significant,
in ordinary  (electrically neutral)  matter one has:
\begin{equation}
V = V({\nu_e}) - V({\nu_\mu}) = V({\nu_e}) - V({\nu_\tau}) = \sqrt{2}
\,G_F \, n_e
\end{equation}
where $G_F$ is the  Fermi constant and $n_e$ is the electron  density.
The effective potential for  $\overline{\nu}$  
is the opposite of   the  $\nu$  one: $V(\overline{\nu}) = - V(\nu)$.
For  neutrinos   traveling  close  to the  Earth  surface 
($\rho \simeq 2.8$~g~cm$^{-3}$, $n_e \simeq 8.4 \times
10^{23}$~cm$^{-3}$)  $V \simeq  1.06 \times 10^{-13}$~eV,
that corresponds to a length
\begin{equation}
 V^{-1}  \simeq 1850 ~\left ( { 2.8~{\rm g~cm}^{-3} \over \rho}
\right )\; \left ( {0.5 \over Y_e} \right )~ {\rm Km}.
\end{equation}
($Y_e$ is the number of  electrons per nucleon).
The effective Hamiltonian for $\nu$'s or
$\overline{\nu}$'s  propagating in matter 
can then  be written:
\begin{equation}
{\cal H}(\nu)    = {\cal H}_0    + {\cal H}_{\rm m},~~~~~~~~~~~~~~~~~ 
{\cal H}(\overline{\nu}) = {\cal H}_0^*  - {\cal H}_{\rm m}
\end{equation}
 as the sum  of the vacuum  hamiltonian:
\begin{equation}
{\cal H}_0 = 
{1 \over 2 E_\nu} \,U~ 
{\rm diag} [m_1^2,  m_{2}^2,  m_{3}^2]  ~ U^\dagger
\end{equation}
 and  a matter    term  that in the flavor  basis 
(neglecting   a  term proportional to the unit  matrix)
has the form:
\begin{equation}
({\cal H}_{\rm m})_{\alpha \beta} = V\;\delta_{\alpha e} \, \delta_{\beta e}
\end{equation}
The effective  Hamiltonian in matter  can  be   diagonalized
to obtain   effective  squared masses  values  and  an  effective mixing
matrix  in matter
that will in general be different for $\nu$ and $\overline{\nu}$.
The matrices  $U_{\rm m}^{\nu}$,  $U_{\rm m}^{\overline{\nu}}$,  
can be  parametrized
with the form (\ref{eq:mixingmatrix})  obtaining   the parameters:
$\theta_{12}^{m,\nu}$,
$\theta_{13}^{m,\nu}$,
$\theta_{23}^{m,\nu}$
and  $\delta^{m,\nu}$, and  similarly for $\overline{\nu}$. 
The   solution of this  problem involves   a cubic  equation
and can  be solved  analytically \cite{ZG}
to obtain  the effective parameters  as a function of the 
product $V~E_\nu$;   however 
the solution is  sufficiently complex not to be 
particularly illuminating, and is  not repeated here.

One representative example of   the dependence  
of the  effective squared masses and  mixing parameters  in matter
on the product $V E_\nu$
is shown  in figures~\ref{fig:mass}, \ref{fig:mixing-parameters-1}
and ~\ref{fig:mixing-parameters-2}
(the density $\rho$    is proportional  to the potential $V$
for  a constant value   of the electron fraction  $Y_e = {1 \over 2}$).
Several important features are clearly visible  most
notably the  ``resonance'' for the angle $\theta_{13}$
at $E_\nu \simeq |\Delta m^2_{23}|/(2 V)$.
More discussion is  contained in  appendix C.

\section{Oscillation probabilities in  vacuum}
\label{sec:osc-vac}
The calculation  of the oscillation probabilities in vacuum is 
a well  known problem,  that is  briefly   outlined here.
The evolution equation for a  neutrino  state is:
\begin{equation}
i{ {\rm d} \over {\rm d}x} \nu_\alpha = {\cal H}_0\,
  \nu_\alpha  
=  \left [ \lambda {\mathbf{1}} +  {1 \over 2 E_\nu} \,U~ 
{\rm diag} [m_1^2,  m_{2}^2,  m_{3}^2]  ~ U^\dagger \right ]~ \nu_\alpha
\label{eq:evoleq}
\end{equation}
where 
 the Hamiltonian ${\cal H}_0$ has  been written
separating a
term proportional to the  unit matrix 
that can be  dropped
because it is irrelevant to the transition
probabilities,
  $m_1$,  $m_2$ and $m_3$ are  the   $\nu$ mass eigenvalues,
and $U$ is the mixing matrix.
The solution of this equation for  $x = L$ is:
\begin{eqnarray}
\nu_\beta(L) = \left \{  U~ \exp \left ( -i
{L \over 2E_\nu}
{\rm diag}
[m_1^2, m_{2}^2, m_{3}^2] \right  )~ 
U^\dagger \right \}_{\beta \alpha} \nu_\alpha(0).
\end{eqnarray}
The  probability  for the  $\nu_\alpha \rightarrow \nu_\beta$ 
transition ($\alpha, \beta = e, \mu, \tau)$ 
 is then:
\begin{eqnarray}
P(\nu_\alpha \rightarrow \nu_\beta; E_\nu, L) &=&
\left | \sum_{j,k} U_{\beta j} \left \{ e^{ -i
{L \over 2E_\nu} }
{\rm diag}
[m_1^2, m_{2}^2, m_{3}^2] \right \}_{jk}~ 
U^*_{\alpha k} \right |^2\\
&=&\sum_{j,k} U_{\beta j}U^*_{\beta k}U^*_{\alpha j}U_{\alpha k}
\exp \left \{-i (m^2_{j} - m^2_{k}) {L \over 2E_\nu} \right \}.
\end{eqnarray}
This expression can  be  expanded  more explicitely
(here  and in all of  the following we will always consider
only non--diagonal  transitions  $\alpha \ne \beta$,
this is no loss of  generality, since the  survival
probabilities can be obtained  from unitarity) as:
\begin{eqnarray}
P(\nu_\alpha \to \nu_\beta) & = &
{A^{12}_{\alpha\beta} \over 2}  ~  [1 - \cos  \Delta_{12} ]  + 
{A^{23}_{\alpha\beta} \over 2} ~  [1 - \cos  \Delta_{23} ]  + 
{A^{13}_{\alpha\beta} \over 2} ~  [1 - \cos  \Delta_{13} ]  +  \nonumber \\
&~~ & \pm  2\, J  ~ [ \sin  \Delta_{12} +  \sin  \Delta_{23} 
-  \sin  \Delta_{13}]
\label{eq:prob-vacuum}
\end{eqnarray}
where: 
\begin{equation}
\Delta_{jk} = {\Delta m^2_{jk} \, L \over 2\,E_\nu} = 
{(m_k^2 - m_j^2) \, L \over 2\,E_\nu}, 
\end{equation}
\begin{equation}
A_{\alpha\beta}^{jk} =  
  -4 ~{\rm Re} [U_{\alpha j}\, U_{\beta j}^*\, U_{\alpha k}^* \, U_{\beta k}]
\end{equation}
and $J$ is  the Jarlskog \cite{Jarlskog} parameter:
\begin{eqnarray}
J & = & J_{e \mu }^{12} =    - {\rm Im}
[U_{e 1}\, U_{\mu 2}^*\, U_{e 2}^* \, U_{\mu 2}]
\label{eq:jarlskog} \\
& = &  c_{13}^2 \, s_{13} \, s_{12} \, c_{12} \,
 s_{23} \, c_{23} \, \sin \delta
\end{eqnarray}
where we  have  also given explicitely the
expression   in  terms of the mixing  parameters  used 
in our  convention. 
The contribution of the first line in equation (\ref{eq:prob-vacuum}) is
symmetric  under a time  reversal  (a  replacement $\alpha
\leftrightarrow \beta$)  or  a $CP$  transformation 
(a  replacement $U \leftrightarrow  U^*$), while the 
contribution of the second  line 
changes  sign both  in case of a $T$ or $CP$  transformation
(remaining   identical for $CPT$  tranformations).
The $+$  ($-$) in  (\ref{eq:prob-vacuum}) is  valid
for  neutrinos   (anti--neutrinos) 
when ($\alpha$,$\beta$)  are in cyclic 
  order, that is:
($\alpha,\beta)=(e,\mu)$, ($\mu$,$\tau$), ($\tau$,$e$).
For   ($\alpha$,$\beta$)  in anti--cyclic
order the sign must be  reversed. 
The term  [$\sin\Delta_{12} + \sin\Delta_{23} - \sin\Delta_{13}$]
can also be  rewritten as the product:
$4 \sin(\Delta_{12}/2)\,\sin(\Delta_{23}/2)\,\sin(\Delta_{13}/2)$.

\subsection{High energy limit}
It is  interesting to consider the  oscillation
probability in the limit:
\begin{equation} 
{\Delta m_{12}^2 L \over 4 \, E_\nu } \ll 1
\label{eq:highenergy}
\end{equation}
that is in the approximation when the oscillations  associated  
 to the longest frequency
cannot develop.
  If $\Delta m^2_{12}$ is in the range suggested 
by the  solar neutrino data, the condition
(\ref{eq:highenergy})  will be satisfied   in
 all   proposed terrestrial  experiments.
Developing   equation  (\ref{eq:prob-vacuum})
in  first order in $\Delta_{12}$  (and using $\Delta_{13} = \Delta_{23} + 
\Delta_{12}$)
one obtains:
\begin{equation}
 P(\nu_\alpha \to \nu_\beta)  = 
{( A_{23}^{\alpha\beta} + A_{13}^{\alpha\beta}) \over 2}
   [1 - \cos  \Delta_{23} ]   
+ {A_{13}^{\alpha\beta}\over 2}  ~  \Delta_{12} ~\sin \Delta_{23}
\pm 2 \, J  ~ \Delta_{12} ~[ 1 - \cos\Delta_{23}]
\label{eq:phigh}
\end{equation}
In this expression the first  term is the dominant one and
 oscillates with  the 
frequency  $\Delta m^2_{23}/(2 \, E_\nu)$; the other two  terms 
are corrections    proportional to
$\Delta_{12} = \Delta m^2_{12} \, L/(2 E_\nu)$.
The last  term in  (\ref{eq:phigh}) is of
great interest
because it  describes $CP$ and $T$ violations
effects. These  effects  oscillate
with the same  frequency as the leading term, and 
therefore for the  
detection it is convenient
to choose   $L$ and $E_\nu$ 
so that 
$|\Delta m_{23}^2| L/(2 E_\nu) = (2n  + 1)\,\pi$  with  $n$ an
 integer. When this  condition is satisfied 
the $CP$ and $T$ violation effects  have  a  maximum.
The  amplitude of the oscillations of the $CP$ and $T$ violating
 effects   is    proportional to the 
Jarlskog  parameter, and to $\Delta m^2_{12}$, 
therefore  the possibility of the  detection of
$CP$ and $T$ violation  effects is  possible 
only  if three conditions  are satisfied:
\begin{enumerate}
\item $\Delta m^2_{12}$ is  sufficiently large.
\item $\sin 2 \theta_{12} = 2 \,c_{12} \, s_{12}$ is  large.
\item $\theta_{13}$ is also large.
\end{enumerate}
The first two conditions  are  satisfied  only if the 
explanation to the solar neutrino  problem is the LMA solution.
Note also that the amplitude of the $CP$ and $T$ violation effects 
also  grows  as $\propto L/E_\nu$.

Examples of the  oscillation  probabilities
$P(\nu_e \to\nu_\mu)$  and  $P(\overline{\nu}_e \to  \overline{\nu}_\mu)$ 
in the regime  discussed here can be seen 
in fig.~\ref{fig:prob_730}  and in the top panels
of fig.~\ref{fig:prob_3000}  and~\ref{fig:prob_7000}
that  illustrate the  qualitative features 
discussed above.

\subsection{Very high  energy limit}
For very large   $E_\nu$  (keeping  $L$  fixed),
also the  fast oscillations   connected with the
larger $|\Delta m^2_{23}|$   cannot  fully  develop,
it is  then    possible to 
rewrite   expression (\ref{eq:prob-vacuum})
as  a power series:
\begin{equation}
 P(\nu_\alpha \to \nu_\beta) 
= 
A_{\alpha \beta} 
\left  (  {\Delta m^2_{23} \, L \over 4\, E_\nu}  \right )^2 
+B_{\alpha \beta} 
\left  (  {\Delta m^2_{23} \, L \over 4\, E_\nu}  \right )^3 
+ \ldots 
\label{eq:vacuum-vh}
\end{equation}
where the  constants $A_{\alpha\beta}$ and  $B_{\alpha\beta}$ are:
\begin{equation}
A_{\alpha\beta} =  A_{\alpha \beta}^{12} \; x_{12}^2  +
A_{\alpha \beta}^{13} \; (1+ x_{12})^2 +
A_{\alpha \beta}^{23}
\label{eq:aa}
\end{equation}
and
\begin{equation}
B_{\alpha\beta} = \pm 8\, J \, x_{12} \, (1 + x_{12}) 
\label{eq:bb}
\end{equation}
Note that  $A_{\alpha\beta}$  is  symmetric  for $CP$ and $T$
transformations
while  $B_{\alpha\beta}$ is anti--symmetric.

\section{Oscillation probabilities in matter}
In this  section we  will  develop some expressions for the
oscillation probabilities  in matter with  constant density.
The density along  the trajectory of a neutrino  traveling
inside  the Earth  will  change  slowly, and for the interpretation
of real  data it will be necessary to integrate numerically
the flavor evolution equation taking into account these  variations,
however  it is a good approximation, sufficient for the 
purposes of this  discussion,  to  consider the density
constant for all  trajectories that do not cross the
mantle--core boundary, that is all  trajectories that  have
$L \aprle 1.06\times 10^4$~Km.
In the approximation of constant density the problem
of  calculating the oscillation probabilities  
is elementary,
 in fact one can simply  use
the expressions  developed in the previous  section 
with the replacements $U \to U_{\rm m}$  and $m_j^2 \to  M_j^2$,
where  $U_{\rm m}$ and  $M_j^2$ the  effective  mixing matrix  and
squared mass eigenvalues in matter, that can be easily 
calculated  as  a function  of the    parameter  $2 \,V E_\nu = 2\,\sqrt{2}
G_Fn_e E_\nu$. 
The limit of this  
 approach is  that the expressions
for the effective  mixing  parameters in matter  are
complicated, and the results are not transparent.

In order to gain  understanding, 
have   calculated  an expression for    the oscillation
probabilities that is  valid  in the limit  of large $E_\nu$,
or more rigorously for $y = |\Delta m^2_{23}| L/(4 E_\nu) < 1$.
In this  situation it is  interesting to write down the oscillation
probability as a power series in $y$, generalising equation 
(\ref{eq:vacuum-vh}).
The  first three  terms   of this expansion
are:
\begin {eqnarray}
P (\nu_\alpha \to \nu_\beta) & = &
A_{\alpha \beta} \; \left( {\Delta m^2_{23} \, L \over 4\, E_\nu }
\right ) ^2 ~
\left [ \left ( {2 \over L\, V} \right )^2 \; 
\sin^2 \left ( { L\, V \over 2} \right ) \right ] 
\nonumber \\
 & + & B_{\alpha \beta} \; \left ( {\Delta m^2_{23} \, L\over 4  E_\nu}
\right)^3 
 \left [ \left ( {2 \over L\, V} \right )^2 \;
 \sin^2 \left ( { L\, V \over 2} \right ) \right ] 
\label{eq:prob-asymptotic} \\
& + & C_{\alpha \beta} \; 
\left  ( {\Delta m^2_{23} \, L\over 4 \, E_\nu }  \right)^3
~L ~V ~
\left \{ 
{48 \over (L\, V)^4 } 
\left [ 
\sin^2 \left ( { L\, V \over 2} \right ) 
- \left ( {L\, V \over 4} \right ) \;
\sin ( L \, V) 
  \right ]
  \right \}  \nonumber \\
& + &  \ldots 
 \nonumber
\end{eqnarray}
The  quantities  $A_{\alpha \beta}$,
$B_{\alpha \beta}$ and $C_{\alpha \beta}$ are  adimensional 
constants  that depend only on the  
ratio  $x_{12} = \Delta m^2_{12}/\Delta m^2_{23}$ and the
four neutrino mixing parameters.
They have the important  symmetry  properties:
\begin{eqnarray}
& & A_{\alpha \beta} =  
+A_{\beta \alpha } = 
+A_{\overline{\alpha} \overline{ \beta}} = 
+A_{\overline{ \beta} \overline{\alpha}}
\\
& & B_{\alpha \beta} =  
- B_{\beta \alpha } = 
- B_{\overline{\alpha} \overline{ \beta}} = 
+ B_{\overline{ \beta} \overline{\alpha}}
\\
& &C_{\alpha \beta} =  
+C_{\beta \alpha } = 
-C_{\overline{\alpha} \overline{ \beta}} = 
-C_{\overline{ \beta} \overline{\alpha}}
\end{eqnarray}
Equation (\ref{eq:prob-asymptotic}) is the  main  result  of this
paper, it is derived    in appendix A.  Note that 
the oscillation probability  (in the approximation of high $E_\nu$)
has been  written as the sum of  three contributions
\begin{enumerate}
\item A leading  contribution  $\propto E_\nu^{-2}$ 
that is invariant for $CP$ or $T$ transformation.
\item A $T$ and $CP$  violating contribution $\propto E_\nu^{-3}$ 
that changes sign  both  for a time  reversal and a $CP$ 
transformation
(and is  therefore   invariant for a
$CPT$ transformation)
\item  A third contribution also  $\propto E_\nu^{-3}$ 
that is  induced by matter effects. This contribution 
is symmetric for a time  reversal
but changes sign exchanging  $\nu$ with $\overline{\nu}$.
\end{enumerate}

The adimensional constants 
$A_{\alpha \beta}$,
$B_{\alpha \beta}$, and
$C_{\alpha \beta}$,
can be calculated   from the neutrino masses and mixing:
It is  remarkable  that the  first two constants are {\em  identical}
to the coefficients   in  the vacuum case (equations (\ref{eq:aa}) and
(\ref{eq:bb})). They can  be  written as:
\begin{equation}
A_{\beta \alpha} = 
({\cal H}_0)_{\alpha \beta} \;({\cal H}_0)^*_{\alpha \beta} ~
\varepsilon^{-2}
\label{eq:A}
\end{equation}
\begin{equation}
B_{\beta \alpha} =   {\rm Im} [({\cal H}_0)^*_{\alpha \beta} \;
({\cal H}_0^2)_{\alpha \beta}] ~
\varepsilon^{-3}
\label{eq:B}
\end{equation}
\begin{equation}
C_{\beta \alpha} =   \pm {1\over 6} {\rm Re}
 \{ ({\cal H}_0)^*_{\alpha \beta} \;
[2\, ({\cal H}_0)_{\alpha e} \, ({\cal H}_0)_{e \beta} - 
({\cal H}^2_0)_{\alpha \beta}
\,(\delta_{\alpha e} + \delta_{\beta e}) ] \} ~
\varepsilon^{-3}
\label{eq:C}
\end{equation}
here    ${\cal H}_0$ is the free  Hamiltonian (equation (\ref{eq:evoleq})),
$\varepsilon = \Delta m^2_{23}/(4 E_\nu)$, and  the $\pm$ sign  refers
to  $\nu$ ($\overline{\overline{\nu}}$).
It can  be checked that  adding to the Hamiltonian a term
proportional 
to the unit matrix  the coefficients do not  change.

The three contributions   to the  oscillation probability have  different
dependences on the  neutrino pathlength $L$.
These  dependencies  are also simple power  laws  when  $L$ is
shorter  than $\sim 2\, V^{-1}$  (in practice  when $L$ is  shorter than 
$\sim  1500$~Km.
Developing  equation (\ref{eq:prob-asymptotic})
for small 
$VL$  one obtains:
\begin {eqnarray}
P (\nu_\alpha \to \nu_\beta) & \simeq &
A_{\alpha \beta} \; \left( {\Delta m^2_{23} \over 4\, E_\nu }
\right ) ^2 ~ L^2 
~ \left [ 1 - { (V \, L)^2 \over 12}  + \ldots \right ]
\nonumber \\
 & + & B_{\alpha \beta} \; \left ( {\Delta m^2_{23}\over 4  E_\nu}
\right)^3  ~L^3
~ \left [ 1 - { (V \, L)^2 \over 12}  + \ldots \right ]
\label{eq:prob-asymptotic1} \\
& + & C_{\alpha \beta} \; 
\left  ( {\Delta m^2_{23} \over 4 \, E_\nu }  \right)^3
~L^4 ~V ~
~ \left [ 1 - { (V \, L)^2 \over 15}  + \ldots \right ]
  \nonumber
\end{eqnarray}
The three contributions  to the oscillation probability
have  dependences
$\propto L^2$ for the leading term,
$\propto L^3$ for the $CP$ and $T$ violation effects,
and
$\propto L^4$ for the matter induced  effects.

To clarify the simple meaning of this equation
let us consider  an experiment with a fixed  baseline $L$.
The oscillation probabilities
for the 4 reactions  connected   by $CP$ and $T$ transformations can
be written   as a  power  series in $E_\nu^{-1}$:
\begin{eqnarray}
P(\nu_\alpha \to \nu_\beta) & =  & {A \over E_\nu^2 }
+  {B \over E_\nu^3 }  +  {C \over E_\nu^3 }  
+ \ldots
\\
P(\nu_\beta \to \nu_\alpha) & =  & {A \over E_\nu^2 }
-  {B \over E_\nu^3 }  +  {C \over E_\nu^3 }  
+ \ldots
\\
P(\overline{\nu}_\alpha \to \overline{\nu}_\beta) & =  & {A \over E_\nu^2 }
-  {B \over E_\nu^3 }  -  {C \over E_\nu^3 }  
+ \ldots
\\
P(\overline{\nu}_\beta \to \overline{\nu}_\alpha) & =  & {A \over E_\nu^2 }
+  {B \over E_\nu^3 }  -  {C \over E_\nu^3 }  
+ \ldots
\end{eqnarray}  
These probabilities have a leading term that is  equal  for  all
four channels, and two  next order terms  one (proportional to  $B$)
 that is due to  the fundamental  $CP$ violation   effects,  and one
(proportional to $C$) that  is the result  of  the matter effects
on the neutrino  propagation.
If one could  excavate  a tunnel along the neutrino path
(to  obtain  vacuum  oscillations)
the  $C$  term   in  the probability would vanish, while the
$A$ and $B$   would   be modified.  For $L \aprle 3000$~Km  the
value of  $A$ and $B$ are  approximately  equal  in matter and in
vacuum.

In an ideal  experimental  program one   could measure 
all four  transitions, and determine separately
the coefficient $A$, $B$ and $C$,  however, if  we  only   two channels
are  experimentally accessible  
(for example
$\nu_e \to  \nu_\mu$  and $\overline{\nu}_e \to \overline{\nu}_\mu$)
a  single  (even  ideal) experiment cannot  disentangle the 
matter effects  from   $CP$  violations      because they have the
same  functional   dependence.
There  are  several   possible  strategies to 
solve this  ambiguity.  One  solution is to perform  two experiments 
with  different  baselines.
The coefficients $B$ and $C$  have different 
dependences on $L$
(for  $ L$ smaller  than $2 V^{-1}$  the  dependences  are approximately
$\propto  L^3$ and  $\propto L^4$ respectively), and a comparison
of the oscillation  rates of the two experiments allow  in principle
to separate the two effects.

It can appear  surprising that
for large $E_\nu$  the oscillation probabilities
in matter  are so  similar
to the vacuum case, since  the mixing parameters
differ  dramatically  from the vacuum case.
This is  the result of some  remarkable cancellations.
For  example   it is possible to show 
that the combination   of parameters 
\begin{equation}
{\cal F} =  J~ \Delta m^2_{12} ~ \Delta m^2_{23} ~ \Delta m^2_{13} 
= c_{13}^2 \, s_{13} \,  
  c_{12} \, s_{12} \,  c_{23} \, s_{23} \, \sin \delta ~  
 \Delta m^2_{12} ~ \Delta m^2_{23} ~ \Delta m^2_{13} 
\label{eq:fcomb}
\end{equation}
 is {\em  independent}  from the matter  effects,  that is 
\begin{equation}
{\cal F}_{{\rm mat},\nu}
= {\cal F}_{{\rm mat},\overline{\nu}}
= {\cal F}_{\rm vacuum}
\label{eq:cancel}
\end{equation}
This  has  the important 
consequence that the $CP$ violating term  of the oscillation probability
is also  independent from the  matter  effects if the $\nu$ pathlength 
$L$ is short
with respect to  the three  vacuum oscillation lengths 
($4 \pi/|\Delta m^2_{jk}|$) and the matter  length $2\pi/V$.
In fact:
\begin{eqnarray}
\Delta P_{CP} & = &2 \, J~
\left [
 \sin \left ( { \Delta m^2_{12} \, L \over 4 E_\nu} \right ) \;
 \sin \left ( { \Delta m^2_{23} \, L \over 4 E_\nu} \right ) \;
 \sin \left ( { \Delta m^2_{13} \, L \over 4 E_\nu} \right )  \right ]
\nonumber  \\
& \simeq  &  {1 \over 32} \;{ L^3 \over E_\nu^3 } ~J ~
 \Delta m^2_{12} ~ \Delta m^2_{23} ~ \Delta m^2_{13} 
=
 {1 \over 32} \;{ L^3 \over E_\nu^3 } ~{\cal F}
\end{eqnarray}
Similarly the  combinations:
\begin{equation}
{\cal G}_{\alpha\beta} = 
A_{\alpha\beta}^{12} ~(\Delta m^2_{12})^2 +  
A_{\alpha\beta}^{13} ~(\Delta m^2_{13})^2 +  
A_{\alpha\beta}^{23} ~(\Delta m^2_{23})^2 +  
\end{equation}
where $A_{\alpha \beta}^{jk} = 
  -4 ~{\rm Re} [U_{\alpha j}\, U_{\beta j}^*\, U_{\alpha k}^* \, U_{\beta k}]$
are also  independent  from  the matter effects.

The existence of these cancellations  has  been discussed in detail
in \cite{lipari-lbl}  in the one mass--scale  approximation  
(the limit $\Delta m^2_{12}\to 0$). The general  demonstration
is  technically  more   demanding,  however the results can be 
readily  verified   numerically.
A   simpler and much   more interesting  demonstration of these  results
can be obtained  not by the brute  force  approach of 
comparing   the  product of   explicit expressions  the parameters,
but   using the power series expansion  outlined in appendix A.

It can be however interesting to see the ``magic'' of the cancellation
in (\ref{eq:cancel})  in an explicit example  
that is  worked out in appendix  C.

\section{Discussion}

Plots of the oscillation probabilities 
for  the transitions $\nu_e \to \nu_\mu$
and $\overline{\nu}_e \to \overline{\nu}_\mu$
as a function of $E_\nu$ for three  different  values of the
pathlength: $L= 730$, 3000 and 7000~Km
 in   vacuum and in matter
are  shown in  
fig.~\ref{fig:prob_730},~\ref{fig:prob_3000}  and~\ref{fig:prob_7000}. 
To   illustrate  more clearly  the behaviour of the 
oscillation probability  for high $E_\nu$,
in fig.~\ref{fig:eprob0_730},~\ref{fig:eprobm_730} and~\ref{fig:eprob_3000},
 we show   plots of the oscillation  probability
multiplied  by $E_\nu^2$.
To compute the matter effects we have used the 
exact expression  assuming  a homogeneous medium with constant 
density  along the $\nu$ path. 
Since trajectories 
with longer $L$   reach   deeper inside the Earth  where the density
is larger, the estimated average density
is  a function of the
 pathlength: for $L =730$, 3000 and 7000~Km we have used
$\rho = 2.84$, 3.31  and  4.12~g~cm$^{-3}$,  
always assuming an electron fraction $Y_e = 0.5$.
These  examples  exhibit  a number of   striking  features
that  we will discuss   in the following.

\subsection{The measurement of $\theta_{13}$}
The focus of  this  work is  on the measurement  of $CP$ violation  effects, 
however  here are  included  some
comments about the measurement of  $\theta_{13}$.
This  measurement is significantly  easier, 
and the  ``optimization'' of an experimental program
much less ambiguous  than for the   measurement  of $CP$  violation  effects.
It is important to  stress the point that the 
``optimum'' choice of  $L$ and $E_\mu$  for  this  measurement
will  not in general   coincide 
with the  optimum choice  for $CP$  violation  studies.

The key point for this  measurement is 
the   existence
of  a  range of   $E_\nu$  where the probabilities
for   the $\nu_e \leftrightarrow \nu_\mu$  and
$\nu_e \leftrightarrow \nu_\tau$  
are enhanced   because of  matter effects.
This  enhancement, clearly  visible  in  fig.~\ref{fig:prob_3000}
and~\ref{fig:prob_7000},
is  present for  $\nu$'s   if  $\Delta m^2_{23} > 0$ and   for  
$\overline{\nu}$'s
if $\Delta m^2_{23} < 0$, therefore  evidence  of  a non vanishing 
$\theta_{13}$  should also   determine the sign  of  $\Delta m^2_{23}$
\cite{lipari-lbl}.
The enhancement  of the transition probability is related to the 
existence  of an MSW  resonance  for the angle
$\theta_{13}$.
The  effective  angle  $\theta_{13}^{\rm m}$   becomes  in fact
 ${\pi\over 4}$ at an energy  $E_{\rm res}$:
\begin{equation}
 E_{\rm res} \simeq {|\Delta m_{23}^2| \over 2\, V}~ \cos 2 \theta_{13}
\simeq 14.1~ \left ( {|\Delta m^2_{23}| \over 3 \times 10^{-3}~{\rm eV}^2}
\right ) ~ \left ( { 2.8~{\rm g~cm}^{-3} \over \rho} \right )
~{\rm GeV} 
\end{equation}
The position of the enhancement of the oscillation
however does {\em not} 
coincide  with the  resonance  energy but  it is at a  lower
energy $E_{\rm peak} < E_{\rm res}$.
There are three essential points  about the 
matter enhancement that  should be  stressed:
\begin{itemize}
\item  The position  of   the  enhancement   (that is the
value of  $E_{\rm peak}$) 
is  determined
to a good approximation,  for  small $\theta_{13}$, 
 only  by $|\Delta m^2_{23}|$  and the pathlength $L$.
The  value of $E_{\rm peak}$  can be  easily  calculated  exactly,
an  approximate  formula   that  describes reasonably well
the $L$ dependence is:
\begin{equation}
E_{\rm  peak} \simeq { |\Delta m^2_{23}|\; L \over 2\pi + 2 V L}
\label{eq:s13_E}
\end{equation}
note that for small  $L$  this  coincides   with the   highest energy where 
the vacuum oscillation  probability has a maximum
 ($E^* = |\Delta m^2_{23}| L/2\pi$), while  aymptotically
(for  $L\to \infty$)    $E_{\rm  peak} \to E_{\rm res}$.

\item The    size of the  enhancement  of the   
oscillation probability, in good approximation 
(for small  $\theta_{13}$)  depends {\em only} 
on the pathlength $L$. In  reasonably good  approximation
\begin{equation}
{P_{\rm peak} \over   P_{\rm vacuum}} 
\simeq  \left ( 1 - 
{ V  L \over  V  L + \pi} \right )^{-2} \simeq
1 + {2 \over \pi} \, V L + {2 \over \pi} \,(V L)^2 + \ldots
\label{eq:s13_P}
\end{equation}

\item The  width  of the enhancement region
scales  linearly with $|\Delta m^2_{23}|$.
\end{itemize}
Therefore, knowing $\Delta m^2_{23}$,  and  given the   pathlength
$L$ of an  experiment, we  know   a priori 
where (at what $E_\nu$)  the 
enhancement will be  present, and also how large
it will be.
The value of the probability    in the  enhancement
region  is  proportional to $\sin^2 2 \theta_{13}$
that is   unknown, 
 and  therefore  is not predictable  (at least without a
successful theory of the  $\nu$ mixing).

To illustrate these points
in fig.~\ref{fig:s13}  we  plot 
the value of  $E_{\rm peak}$ (top panel)  
and the enhancement  factor 
($P_{\rm peak}/P_{\rm vac}$) 
for  $\Delta m^2_{23} = 3 \times 10^{-3}~{\rm eV}^2$
  and  several  values of $\theta_{13}$
(the values of the other parameters  are to a good approximation
not important).
It can be seen that when   $\theta_{13}$ is small the curves    are
independent  from its  value.

These results can be easily understood qualitatively,  
since for this  purpose 
it is sufficient to approximate
   $\Delta m^2_{12} \simeq 0$. In this  approximation the
oscillation probabilities involving electron
neutrinos are proportional to the two  flavor formulae
(see appendix B).
The  oscillation probabilities for the transitions
$\nu_\mu\to \nu_e$ are then  given by:
\begin{equation}
P_{\nu_e \leftrightarrow \nu_\mu} = {s^2 \sin^2 \theta_{23}
~ \over s^2 + c^2\,(1 \mp E/E_{\rm res})^2 } ~
\sin^2 \left [ {\Delta m^2 \,L \over 4 \,E_\nu} \;
\sqrt{s^2 + c^2\, (1 \mp E/E_{\rm res})} \right ]
\end{equation}
(where we have used the  shorthand notation $s = \sin 2 \theta_{13}$,
$c = \cos 2 \theta_{13}$).
It is  trivial  to obtain   exactly the
energy $E_\nu$   of the absolute  maximum
of this   probability and its  value.
The  position of the maximum  in general does  not correspond to the 
the  resonant energy, when the first factor   is    largest,
because at the resonance the   oscillation length 
becomes  very long:
\begin{equation}
\lambda_{\rm res} 
= {\lambda_{\rm vacuum} (E_{\rm res})  \over \sin 2 \theta_{13}}
\end{equation}
(where $\lambda_{\rm vac} = 4\pi E_\nu/|\Delta m^2_{23}|$
is the  vacuum  oscillation length)
and  the oscillations  do not have  time to develop.
The maximum     is  found at  a lower  energy,  where the
mixing parameter is  smaller  but the phase of the oscillations
is  close to ${\pi \over 2}$.

A  detailed  analysis of the position of the  ``peak'' 
in the oscillation probability and the value
of  the probability at the peak shows the existence
of  a    weak dependence
on the other  parameters of the mixing matrix,
$\Delta m^2_{12}$, $\theta_{12}$ and $\delta$.
Therefore  a careful  study of the  shape of the oscillation probability
for the enhanced channel ($\nu_e \to \nu_\mu$ or $\overline{\nu}_e \to \overline{\nu}_\mu$
depending on the sign of $\Delta m^2_{23}$) can  in principle  give information
on $\delta$. This  line of  research is  actively  pursued \cite{yasuda}.
A  fundamental  difficulty is that since    matter  effects are very
large close to the resonance, they are not easy to subtract, and
the uncertainties  of the  matter density profile
along the neutrino path   can be 
reflected  into   effects  on the probability
of  the same size as the $CP$ violation effects.

\subsection{The ``vacuum  mimicking'' region}
Looking at  fig.~\ref{fig:prob_730}   one  can note   a remarkable
feature,
namely the fact  that the oscillation probabilities for
$L = 730$~Km and $E_\nu \aprle 0.5$~GeV  are approximately independent
from  the presence of matter. 
This  phenomenon  has  been  observed  before,
in particular  by Minakata  and   Nunokawa \cite{minakata},
who  refer to it as ``vacuum mimicking''. 
At first sight the  closeness of the oscillations in matter and
vacuum for  energies  as  large as 0.5~GeV, traveling in 
ordinary matter may seem  surprising,
since  we can expect, and  indeed it is the case, 
 the effective squared
mass  values and mixing parameters  are modified by the
matter potential, however  these  modifications of
the  oscillation parameters are {\em not} reflected in
the oscillation proabilities 
if the  neutrino  pathlength is    sufficiently short, namely
if  $V \, L  \ll 1$.
In fact it can  be  proved  (see appendix A)  that the  
difference  $\Delta P_{\rm matter} =  P_{\rm matter} - P_{\rm vacuum}$
(for  a homogeneous medium)
can be  expressed as  a power  series in
$V\,L$  and    becomes  negligible  for  $V\,L$ small  even 
no  matter  how  different   from the   vacuum  values are  the 
effective  squared masses and mixing  parameters  that correspond
to  the product $V \,E_\nu$.
parameters. 

This  is  illustrated in  fig.~\ref{fig:regions}
that  describes    different  regions in the   plane
$(L,E_\nu)$     where the oscillation probabilities
for the $\nu_e \leftrightarrow \nu_{\mu,\tau}$  transitions 
have  different qualitative  properties.
The  figure is constructed   for  $\Delta m^2_{23} = 3 \times 10^{-3}$~eV$^2$,
$\Delta m^2_{12} = 7 \times 10^{-5}$~eV$^2$, and for the 
potential  $V = 1.06 \times  10^{-3}$~eV  (that  corresponds to 
the Earth's  crust).
\begin{enumerate}
\item  The effects  of  matter   on the oscillation probability
can be   significant  only if  $L$  is   close  or  larger
than $2 V^{-1}$  that is  close  or to  the right of 
the line  labeled  $a$.
For growing  $L$ the
matter  effects  become more and  more important
(compare 
fig.~\ref{fig:prob_730},~\ref{fig:prob_3000} and~\ref{fig:prob_7000})
and  a  more and more serious   background for 
$CP$ violation studies.

\item Curve  $b$  
   is  defined   by  $E_\nu = |\Delta m^2_{23}|\, L/(2\pi)$, and
indicates  the position of the highest energy maximum  of the
transition probabilities,  above this  line they
decrease  monotonically  without  further oscillations.

\item Curve $c$    is  defined  by the relation
 $E_\nu = \Delta m^2_{12} \,L/(2\pi)$  and gives  the  highest energy where the
slow  (``solar'')  vacuum oscillations   have  a  maximum.

\item Matter effects  are  particularly
spectacular   for   $E_\nu \simeq  |\Delta m^2_{23}|/(2V)$
(curve $d$)  when  $\theta_{13}$  undergoes an MSW resonance.

\item The matter  effects   have  negliglible 
effects  on the   effective squared masses and  mixing 
(and therefore no  effect   on the   transition probabilities)
if $E_\nu  << \Delta m^2_{12}/(2 V)$  (curve  $e$).
For  $\theta_{12} \ne {\pi \over 4}$  this  corresponds also
(adding a factor  $\cos 2 \theta_{12}$) to the location of the
MSW resonance for  the  angle  $\theta_{12}$.

\item The  ``high energy region''   that is the  main focus of this   work
corresponds to the region  above the thick dot--dashed  lines,
when the  probability is not oscillating any more.

\item The ``vacuum mimicking region''     corresponds to the
short   pathlength   (significantly  shorter  than  $2 \, V^{-1} \simeq
3700$~Km.  In  order  to access $CP$ ($T$)  violation effects  where
they are  large, the energy must also  be  small  enough to  see
the development of   several oscillations.
This  qualitatively corresponds  to the region   delimited  by the 
thick dashed line.

\item Finally  the thick  curve  labeled $A$ indicate the energy where the 
MSW  enhancement   (or suppression)  of  the  probability
is    most  important  (correspond to $E_{\rm peak}$  discussed in
the previous  subsection.
The  line is  plotted  only when the enhancement of the
probability is  larger  than 20\%.
The  enhancement   becomes  more  and  more  important  with  growing  $L$
(see equation (\ref{eq:s13_P}).
The best place  to  search for  a non  vanishing $\theta_{13}$   is close
to this  line.
\end{enumerate}

\subsection{High Energy Neutrinos}
We will now  discuss in more detail  the behaviour of the oscillation
probabilities for high $E_\nu$, when they are 
monotonically decreasing.
Some  important features 
of the oscillation probability in vacuum
are  illustrated  in fig.~\ref{fig:eprob0_730}
thas  shows the   product $P_{\nu_e \to \nu_\mu} \times E_\nu^2$
   plotted as a function
of $E_\nu$ for  a  fixed  value  $L = 730$~Km:
\begin{enumerate}
\item For  large $E_\nu$   the oscillation probability
is well approximated  with the form $A/E_\nu^2$.
\item The  value  of the  constant $A$
(keeping all other parameters  fixed)
depends  on the value of the  $\cos \delta$
(this is  a crucial  remark).
\item  The $CP$ violation  effects  (present when   $\sin\delta \ne 0$)
have an energy dependence  $\propto E_\nu^{-3}$.
\end{enumerate}
Fig.~\ref{fig:eprobm_730} illustrates how the
oscillation probability is modified by the  matter  effects
(for the same $L = 730$~Km and the same   $\nu$ masses and mixing
as the previous   figure).
The   probabilities  in vacuum and in  matter  
for the longer  pathlength 
$L = 3000$~Km
are shown 
in fig.~\ref{fig:eprob_3000}.
Some  important points are the following:
\begin{enumerate}

\item Matter effects  generate  a $\nu$/$\overline{\nu}$
asymmetry.

\item  The  matter induced asymmetry 
has the same  energy dependence ($\propto E_\nu^{-3}$)
as   the one  generated  by the fundamental  $CP$ violation effects.

\item  The effects 
of  the phase $\delta$  and  of matter 
can both  contribute  to the observable  $\nu/\overline{\nu}$  asymmetry,
either adding or subtracting from each other.

\item   The effects  of   matter 
depend strongly on the pathlength $L$.

\item  The  relative  importance of the matter
induced asymmetry  with respect to the asymmetry  
generated  by   the fundamental  $CP$ violation effects
grows approximately linearly  with $L$
(see equations (\ref{eq:prob-asymptotic}) and (\ref{eq:prob-asymptotic1})),
 and  the  ``background''  of the matter induced asymmetry is a 
 much more serious  problem  at $L = 3000$~Km.

\item  The effect   of the presence of matter on the 
leading order term ($A/E_\nu^2$)  of the   transition probability
also depend on  the pathlength $L$
\begin{itemize}
\item [$-$] For  short $L$  the 
constant in  matter $A_{\rm mat}$  is  equal to the vacuum value $A_{\rm vac}$.
\item [$-$] For longer  $L$  ($L \aprge 2\, V^{-1}$)  the leading  term
of the oscillation probability  is  suppressed 
(see again equations (\ref{eq:prob-asymptotic}));
 however  one has 
$A_{\rm mat} = F(LV) \times A_{\rm vac}$, that  is the 
constant  $A$ is 
proportional to the vacuum value    with a  proportionality 
factor that  depends  only  on the product of
$V \,L$; therefore  a measurement  of 
the leading  term  again  carries  information about $\cos\delta$.
\end{itemize}
\end{enumerate}

\subsection{The leading term in the oscillation probability and $|\delta|$}
The  leading  term  in the oscillation probability
$\sim  A_{\alpha \beta}/E_\nu^{2}$  is  equal  for all  the four transitions
related  by a $CP$ or a $T$  transformation,
however  in principle  its measurement, together with a precise determination
of the other  oscillation parameters, can give
information  about the  phase  $\delta$. 
In fact  most of the   sensitivity  to $\delta$  claimed
by  recent analysis  of the potential of
high energy $\nu$--factories,  is  essentially
 the result of  a high precision measurement of the constant $A$.
Considering  all other parameters as fixed
and $\Delta m^2_{23}$  positive, 
the constant $A_{e\mu}$
is  largest when  $\delta = 0$,  
and decreases  monotonically with growing $|\delta|$, reaching a
minimum value for $|\delta| = \pi$.
Conversely   $A_{e \tau}$ is 
maximum  for $|\delta|=\pi$ 
and minimum for $\delta = 0$.
For $\Delta m^2_{23} < 0$ the   dependence  of
$A_{e \mu}$ and  $A_{e \tau}$ on   $|\delta|$
is reversed. 

It is  simple to   give a qualitative explanation for this behaviour.
The constant  $A_{\alpha\beta}$ can be expressed   as the 
 sum of  three contributions
each one  associated to a squared mass difference:
\begin{equation}
A_{\alpha \beta} = 
A_{\alpha \beta}^{12}  \times x_{12}^2 + 
A_{\alpha \beta}^{13} \times (1 + x_{12})^2 +
A_{\alpha \beta}^{23}  \times 1
\label{eq:acomb}
\end{equation}
In this  expression each factor 
$A_{\alpha \beta}^{jk} = -4{\rm Re} [U_{\alpha j} \, U_{\alpha k}^* \,
U_{\beta j}^* \, U_{\beta k}]$ is  weighted  proportionally to
to the square of  the  relative squared mass difference  $\Delta m^2_{jk}$. 
The general expressions for the different  terms  in our parametrization
of the mixing  matrix are easily calculated
. For the $\nu_e \leftrightarrow \nu_\mu$ transitons  one has:
\begin{equation}
A_{e\mu}^{12}  =
 4 \, s_{12}^2 \, c_{12}^2 \, c_{13}^2 \, (c_{23}^2 - s_{13}^2 \, s_{23}^2)
+ (c_{12}^2 - s_{12}^2) \,s_{13} \, s_{23} \, c_{23} \, \cos \delta
\end{equation}
\begin{equation}
A_{e\mu}^{13}  =
 4 s_{13}^2 \, c_{13}^2 \, s_{23}^2 \, c_{12}^2  +  4 \, s_{12} \,
 c_{12}  \,s_{13} \,
 c_{13}^2 \, s_{23} \, c_{23}
 \, \cos \delta
\end{equation}
\begin{equation}
A_{e\mu}^{23}  =
 4 \, s_{13}^2  \,c_{13}^2 \, s_{23}^2 \, s_{12}^2 -  4 \, s_{12} \,
 c_{12} \, s_{13} \,
c_{13}^2 \, s_{23} \, c_{23}
 \, \cos \delta
\end{equation}
One can see that the term  $A_{e \mu}^{12}$  is  non  vanishing 
also for $s_{13} = 0$,  and in fact is  related to the 
``solar oscillations''.
Note also that  the sum  $A_{e \mu}^{13} + A_{e \mu}^{23}$ 
is independent  from $\delta$ (and from $\theta_{12}$),
however the  two  individual  terms   do  depend on the phase
(and also on  $\theta_{12}$).
For  $\cos \delta \to 1$  the term 
$A_{e \mu}^{13}$   becomes  largest
 while   $A_{e \mu}^{23}$ becomes smallest;
this is  a  consequence  of the fact
that the  overlap  $|\langle \nu_\mu | \nu_1\rangle|$
is largest for $\cos \delta=1$
(see discussion in section~\ref{sec:delta}).
This has  simple but important  consequences
when considering the combination (\ref{eq:acomb}).
When   $x_{12}$ is positive  (that is for $\Delta m^2_{23}$ positive)
one  has   $|\Delta m^2_{13}| > |\Delta m^2_{23}|$
and the contribution $A^{13}$   receives the largest  weight,
 therefore the $e$--$\mu$ oscillations have the highest  probability 
when  $A_{e\mu}^{13}$ is largest  that  is  when
 $\cos\delta=1$.
For  $x_{12}$ negative  ($\Delta m^2_{23} < 0$) 
one  has   $|\Delta m^2_{13}| > |\Delta m^2_{12}|$
and  therefore it is 
the  term $A^{23}$   that   receives  the larger  weight,
therefore the probability is largest when $A_{23}$ is  largest, that is
for $\cos \delta = 0$.
A  similar discussion can be  performed  for $e$--$\tau$ transitions.
In this  case the oscillation probability
is largest when   $\cos \delta = 0$  
for $\Delta m^2_{23}  >0$, 
and  when  $\cos \delta = 1$  
for $\Delta m^2_{23}  < 0$.

For  a simple  illustration  we can   consider  the  
case of  quasi--bimaximal mixing
$(\theta_{23} = \theta_{13} =  {\pi \over 4})$.
Keeping only terms  in first and  second  order in  $s_{13}$ 
one obtains:
\begin{equation}
A_{e \mu, e \tau} = {1 \over 2}\, (1 - s_{13}^2) \, x_{12}^2
 + 2 \, s_{13}^2 
 +  s_{13} \,  \cos \delta  \,[ (1 +x_{12})^2 - 1]
\label{eq:leading-bimax}
\end{equation}
One can recognize   three main contributions that can be considered as
the ``solar contribution'', the ``$\theta_{13}$ contribution'' and
the ``mass splitting effect''.
\begin{enumerate}
\item  The first term  $\propto  x_{12}^2$,  is due to  
the effect  of oscillations involving only the   states  $\nu_1$ and $\nu_2$.
This  contribution  is  non--vanishing 
also for $\theta_{13} = 0$, it is ``garanteed''  to exist,
since it is responsible for the oscillations of solar neutrinos.

\item The second  contribution 
$\propto  s_{13}^2$, can be understood as 
oscillations  between the
state  $\nu_3$ and the quasi--degenerate pairs $\nu_1$--$\nu_2$. 
This  contribution  is  independent from the solution
of the  solar neutrino problem  and can in fact  exist also
for $\Delta m^2_{12} \to 0$, however it 
can  be  arbitrarily small   since
it is  proportional to $s_{13}^2$ for which exists only an upper limit.

\item The third contribution   
is  $\propto  s_{13}\, x_{12} \cos\delta$,  and is  non--vanishing
only  if both $\Delta m^2_{12}$, and $s_{13}$ are non  zero.
It  arises  as the consequence of  the different
weights  for the 
oscillations  ``between'' the pairs of states  $\nu_1$--$\nu_3$,
or $\nu_2$--$\nu_3$ due to the associated
squared  mass  difference.
This is the contribution   that  carries information about the phase  $\delta$.
\end{enumerate}

An important remark 
 is that, for a large 
interval of  values,
when  $s_{13}$   decreases 
it becomes  {\em easier} 
to  measure the contribution of the $\cos\delta$ term,
even if  it becomes  smaller. 
This   happens because  the  leading contribution
is usually  the ``$\theta_{13}$  term''  that  is
$\propto s_{13}^2$   and 
decreases  quadratically
with $s_{13}$,  while  the $\cos\delta$ contribution 
decreases  only linearly with $s_{13}$.
In fact Cervera et al \cite{cervera}
in their analysis of the sensitivity of
 $\nu$--factory experiment 
have found (see  fig.~22 in their work), that the range of
$\Delta m^2_{12}$   where it is possible to
distinguish 
$\delta =0 $ and $\delta = \pi/2$
is  apoproximately constant (with actually a small {\em increase})
when  $\theta_{13}$ becomes smaller,
down to the smallest  angles  they investigated.  This
observation seems a  paradox,  since 
the $CP$ violation effects  
decrease linearly with $s_{13}$.
However  the observation is correct, and 
can  be  easily  understood with the argument  outlined  above.
If the  authors of \cite{cervera}
had studied    the possibility to discriminate  
between $\delta = -{\pi\over 2}$ and 
 $\delta = +{\pi\over 2}$  the allowed region would be
significantly smaller, and  shrink linearly with $\theta_{13}$.
This in  fact reveals
the conceptual  difference between  measuring 
$CP$ violation  effects  and  measuring
the phase $\delta$. 

In summary:
\begin{enumerate}
\item  The  measurement    of the leading term in the oscillation
that is   a   quantity  symmetric  under $CP$ and $T$ transformations
gives information about the value  of $\cos\delta$. 
This  result  does  not allow to  determine the
  sign of  the $CP$ violation effects.

\item A precise measurement of the leading term in the oscillation
probability  is significant as a measurement of
$|\delta|$  only  if  the other oscillations
parameters can  be measured separately with a sufficient accuracy.
The  requirement  on the precision of the measurements 
of the ``solar parameters''
$\theta_{12}$ and $\Delta m^2_{12}$ are  particularly  stringent
(they can be deduced  from eq.~53--57).

\end{enumerate}

\section{Conclusions}
In this  work we have addressed  the question  of the optimum
strategy, that is the  best choice
of $\nu$   energy and
pathlength,  to measure the  two  remaining  completely unkwown parameters
in the neutrino  mixing  matrix \cite{note},
 that is the   angle $\theta_{13}$ and
the phase $\delta$.

For  the measurement of 
$\theta_{13}$  it is   possible to  make a strong case 
for  a high $E_\nu$ and  long $L$ program.
The  oscillation probability for the
$\nu_e \leftrightarrow  \nu_\mu$ (or 
$\overline{\nu}_e \leftrightarrow \overline{\nu}_\mu$  depending
on  the sign  of $\Delta m^2$)   transitions
will be enhanced in a well defined  and  precisely
 predictable  energy range. It is in this
energy range that the search for the  effects of a non vanishing 
$\theta_{13}$  has  the best possibilities.
The maximum  of the oscillation probability 
corresponds to 
an energy $E_{\rm peak}$
that is  proportional to  $|\Delta m^2_{23}|$,
is  only weakly dependent on
 $\Delta m^2_{12}$  and the mixing parameters,
and grows with  increasing $L$ 
in a well defined way (see eq.~\ref{eq:s13_E} and the 
following discussion).
For $|\Delta m^2_{23}| = 3 \times 10^{-3}$~eV$^2$
$E_{\rm peak}$ is approximately 
1.6, 5.2 and  7.4~GeV  for $L = 730$, 3000 and 7000~Km.
The  value of the probability  at  $E_{\rm peak}$ 
can   be predicted as
$P_{\nu_e \leftrightarrow \nu_{\mu}}^{\rm max} \simeq
 \sin^2 2 \theta_{13} \, \sin^2 \theta_{23} \times F$ where $F$ 
is  a matter  enhancement effect
 that to a good approximation   depend  {\em only} on the pathlength 
 $L$  (see eq.~\ref{eq:s13_P}).
For $L = 730$, 3000 and 7000~Km the  enhancement $F$ is
$\sim 1.25$, 2.8 and 10.3.
The best  strategy for a detection of $\theta_{13}$ is therefore
to  use a very long pathlength (since it improves the signal to background)
and design 
a neutrino  beam  with
maximum intensity in the  energy range   where the
probability is  predicted  to have  the maximum.
Note that for this study,   a conventional beam \cite{richter,superbeam}
could be  competitive with a neutrino factory.

The  study of $CP$ and $T$ violation effects is  a fascinating  subject,
and it is  remarkable  
that if two conditions  are satisfied: (i) the  LMA solution is  the 
explanation of the solar  neutrino  problem, and (ii) 
$\theta_{13}$ is    sufficiently large,
these effects are in principle observable with accellator $\nu$ beams,
and the  phase $\delta$  
is  experimentally measurable;
however these  are  extraordinarily difficult tasks, and the best strategy
is not  easily determined.
Very likely in this case 
the very well  controlled and intense  beams  of  a $\nu$  factory
are a uniquely well suited  tool,
however the choice  of $E_\mu$  and  $L$ is  not a  simple decision.
A large $E_\mu$    allows  very high event  rates,
but   results in high energy neutrinos  that have  small
oscillation probabilities  and  for  which  the $CP$ violation effects
are  strongly suppressed ($\propto E_\nu^{-3}$), moreover the
matter effects   become a more  dangerous source of background.

To analyse  quantitatively  if the high rates
of  a high energy neutrino factory 
are  sufficient to extract  information about the phase  $\delta$
we have   analysed in detail the  oscillation probabilities
for high energy neutrinos, expressing the probability
as a power series expansion
in the  adimensional  parameter $ y = |\Delta m^2_{23}| \, L/(4 E_\nu)$.
This  expansion is  useful when $y$ is  less than unity,
that  is  for  $E_\nu$  larger  than  few GeV, even
for the largest possible $L$,
and reveals    several important  features of the probability.
Keeping only the lowest order terms
in the expansion,
the oscillation probabilities is the sum of 
three   contributions: 
\begin{equation}
P = A_0 (\cos \delta) \; {L^2 \over E_\nu^2} 
\pm B_{CP} \; \sin\delta \;{L^3 \over E_\nu^3} 
\pm C_{\rm mat} \;  V \; {L^4 \over E_\nu^3} 
\end{equation}
One  distinguish:
\begin{enumerate}
\item a  leading  order contribution $\propto E_\nu^{-2}$,
that  is  symmetric  under  $CP$   and $T$  transformation

\item  a  $CP$  and  $T$ antisymmetric  contribution of  order $E_\nu^{-3}$,
that depends  linearly on $\sin\delta$,

\item a matter  induced contribution, also  proportional
to  $E_\nu^{-3}$, 
that  is  invariant for a $T$ reversal  transformation,
but changes sign replacing   $\nu$ with $\overline{\nu}$
(or viceversa).
This  term is proportional to the potential $V$ and  vanishes in 
vacuum.
\end{enumerate}
It is important to  note the $L$ dependence 
of the three terms:  longer $L$   enhances the $CP$ violation  effects, but 
enhances  more dramatically  the matter  effects.  

The largest effect of the phase $\delta$ is on the  leading term 
coefficient $A$.  This  coefficient 
can be measured with great precision,  and for  this  purpose
a very high energy is the optimum  solution, however the value  of
$\cos \delta$  extracted  from the  measurement only if
the squared mass differences and   mixing  angles are
determined  (from other  measurements)  with sufficient precision.
The possibility to  obtain the required accuracy
for the  ``solar''  parameters $\theta_{12}$  and  $\Delta m^2_{12}$
is problematic
and  should  be  critically  analysed.

The detection  of   a genuine $CP$ violation effect, is  more
difficult. It requires  first of all
to subtract the matter   induced asymmetry.
This can be  done   having two  experiments
with different  baselines,  
and using the  different  $L$ behaviour of the two contributions,
or  studying the energy dependence of the probability 
 down to lower $E_\nu$.
Note that the  effect on  the event rate
 induced by the fundamental $CP$  violation   is  constant  with 
increasing $E_\mu$, since the increase in  the rate $\propto E_\mu^3$ is
compensated by the suppression in the probability
$\propto  E_\nu^{-3}$  for higher energy neutrinos.
Since  the backgrounds increase with energy, the optimum  solution
for the search of  the asymmetry is not  an arbitrary high energy.

The  authors  of some recent works \cite{cervera,arubbia} on the  
sensitivity of  high energy neutrino  factories
for the determination of
$\delta$  have  neglected  to consider the
entire    interval of  definition  of the phase
 $\delta \in [-\pi,\pi]$, studying only the positive
semi--interval.
It would  be interesting to  see  a  reanalysis  of those  works that 
considers the entire  interval of  definition for  $\delta$.
The  outcome  should  be  that  at least in a significant part
of the parameter space 
 an input   value $\delta_{\rm input}$ 
can be   reconstructed   with  fits of approximately the same
quality as $\delta_{\rm rec} \sim \delta_{\rm input}$  and
$\delta_{\rm rec} \sim  - \delta_{\rm input}$.
This  would  reveal how  much of the  sensitivity is  coming
from the measurement 
of the $CP$  violating part  of the 
oscillation probability, 
and  how much  is  coming from the 
high precision measurement  of the $CP$ conserving   part.

An ambiguity  of sign  in the dermination of $\delta$
leaves ambiguous  also  the 
sign of all $CP$  of $T$ violation effects   in the
neutrino  sector.  This  can obviously be a limitation, for example
for a  discussion  of the relation between these effects 
and the  observed $CP$  asymmetry of the 
present  universe.
More in general,  it should be noted that 
a measurement  of  $\cos \delta$ is not  rigorosly speaking 
a measurement  of $CP$ violations  at all,  but it implies
the existence of $CP$ violations of  predicted  size 
but unknwon   sign.
One point that in my  opinion would require more  attention is the 
following,
the   measurement of  $\cos \delta$  (different 
from the  special values 0 and 1)is equivalent 
to the statement that three flavors,  two squared mass  differences and
three mixing  angles are  insufficient to describe all
observed  results about the   $\nu$ flavor  transitions,
and  that the inclusion of a  new parameter can
reconcile all results.
It is  not   clear if this    interpretation of the data
would be unique. It is  interesting  to discuss  if 
other  mechanisms,  for example   the  introduction of
new  neutrino properties (such as FCNC interactions),  or 
additional  small  mixings (``LSND--like'') 
with  light   sterile  states, could also 
be viable descriptions of  the data.

An experimental  program  with low  energy  neutrinos  and
a short pathlength  has in priciple  some
very attractive features: (i)  the problem of  disentangling
the matter  effects is  much less severe because these effects are
small,  (ii)  a  direct measurement of  $CP$ violation effects
is possible, (iii) the   oscillation probability can  have more
structure, and the  $CP$ violation effects   can be very large
(with $\Delta P_{CP}/P \sim 1$).
Unfortunately, the  experimental difficulties  are
enormous,  because of   poor    focusing,  small cross sections,
and the experimental difficulty of  flavor  determination.   
The  question of  which one  of the two options 
(high energy or low energy) is  more  promising 
for the  measurement   of the phase 
$\delta$  and of   $CP$ violation effects for
leptons  remains in my  view still open, 
and more detailed  studies have still to be  performed.

\vspace {1.5 cm}
\noindent{\bf Acknowlegments} 
This  work was  initiated  by conversations  with Alain Blondel,
Andrea  Donini, Belen Gavela, Michael Lindner and Hisakazu Minakata.
I'm  very grateful  to Maurizio  Lusignoli 
and Andrea  Donini    for  several discussions and  the  kind reading
of  an early version of   this  work.
Lively discussions  with Hisakazu Minakata, Hiroshi Nunokawa and
Osamu Yasuda are gratefully acknowledged.

\newpage

\appendix
\section* {Appendix A: Oscillation probabilities as power series}

\addtocounter{section}{1}

\subsection{Oscillations  in vacuum}
In this  appendix we will show how the $\nu$  oscillation probabilities
both in   vacuum and in  homogeneous matter  can  be  expressed
as  power  series.

We can start  with the vacuum case.  
In this case,  as  discussed in 
section~\ref{sec:osc-vac}, the  oscillation probabilities  can be 
calculated  with simple  analytical  formulae and are 
a function of the ratio  $L/E_\nu$.
It is  useful to  define the  adimensional  quantity $y$:
\begin{equation}
y = {\Delta m^2_{23} \, L \over 4 \,E_\nu}.
\end{equation}
It is  straightforward to   expand  the oscillation
probability as  a power  series in $y$:
\begin{equation}
P_{\alpha \to \beta}  (L, E_\nu)  =
\sum_{n=2}^\infty  c_n^{\alpha \to \beta} ~y^n
\label{eq:expansion-vacuum}
\end{equation}
developing  in a Taylor  series the  trigonometric  functions  in the
analytic  expression (\ref{eq:prob-vacuum}).
The constants $c_n^{\alpha \to \beta}$ 
   are adimensional  quantities  that 
can be written as 
functions  of the squared mass ratio $x_{12} = \Delta m^2_{12}/\Delta m^2_{23}$
and of the 4 mixing  parameters.
The expansion (\ref{eq:expansion-vacuum})  is formally  always valid, it is
of course useful  only when  $y$ is  less  than  unity, that is for
short $\nu$--pathlength  or  high $E_\nu$.
Since the expansion of the $\cos$ ($\sin$)   function
    that  describe the  $CP$  conserving
(violating)  part of the probability  has  only
even  (odd) powers of  their   argument, 
the   coefficients  have the  following
symmetry properties:
\begin{equation}
 \cases { 
c_n^{\alpha \to \beta} =
+c_n^{\beta \to \alpha} =
+c_n^{\overline{\alpha} \to \overline{\beta}} =
+c_n^{\overline{\beta}  \to \overline{\alpha}}
~~~~~          &for $n$~even, \cr
\omit      & \omit \cr
c_n^{\alpha \to \beta} =
-c_n^{\beta \to \alpha} =
-c_n^{\overline {\alpha} \to \overline {\beta}} =
+c_n^{\overline {\beta} \to \overline {\alpha}}
~~~~~          &for $n$~odd, \cr
 }
\end{equation}
Note in particular that the lowest order  (leading)  term of the expansion 
is  exactly  $CP$ conserving,  while the next term
is  $CP$ anti--symmetric.
It follows that  the    determination   in  an experiment   with a fixed
baseline,   that the  flavor  transition probability  in  vacuum 
vanishes at high energy 
with    the form  $P_{\nu_\alpha \to \nu_\beta} \simeq a\,E_\nu^{-2} 
+ b\,E_\nu^{-3}$  with $b \ne 0$, 
would  be  
a proof of  the existence of   $CP$  violations  in the neutrino sector.

It  is  useful to  rederive the  expansion  (\ref{eq:expansion-vacuum}) 
of the oscillation probability  with  a different  method, that can be
more easily extended to the  case   of neutrinos  propagating in 
matter. 
The    $S$  matrix  for    flavor  transition
can  be calculated as:
\begin{equation}
S(\nu_\alpha \to \nu_\beta)  \equiv S_{\beta \alpha}
=  \exp [-i {\cal H}_0 \,L]_{\beta \alpha}.
\end{equation}
where ${\cal H}_0$ is the  free Hamiltonian.
The $S$ matrix  for the  transitions  of  $\overline{\nu}$'s  can  be 
obtained replacing ${\cal H}_0$ with  the complex  conjugate  ${\cal H}_0^*$.
Expanding the exponential  one has:
\begin{equation}
S_{\beta\alpha} = \exp [-i {\cal H}_0\, L]_{\beta \alpha} =  
\delta_{\beta \alpha} 
+ ( -i \, L) ({\cal H}_0)_{\beta \alpha}  + {1 \over 2 !}  ( -i \, L)^2
({\cal H}_0^2)_{\beta\alpha} + \ldots
\end{equation}
The  transition probability can  be obtained  squaring the
corresponding
matrix  element:
\begin{equation}
P_{\alpha \to \beta} =  | S_{\beta \alpha}|^2 
\end{equation}
Collecting all  terms  proportional  to $L^n$  for  each  integer $n$,
 one  can then   obtain the probability 
as a power  series in $L$.
This  actually corresponds to the  powers  series in $y$,  since
 $L$  always  enters in the combination
${\cal H}_0\, L$ or ${\cal H}_0^* \,L$, and the  Hamiltonian can  be written 
(neglecting a term proportional to the unit matrix) as:
\begin{equation}
{\cal H}_0 = 
{\Delta m^2_{23} \over 4 E_\nu} ~
U\; {\rm diag} [-(1 + 2 x_{12}), -1, 1]~U^\dagger 
=
{\Delta m^2_{23} \over 4 E_\nu} ~
{\hat h}
\end{equation} 
and  each  power of ${\cal H}_0$ (or ${\cal H}_0^*$) 
contributes a factor  $\Delta m_{23}^2 /(4 E_\nu)$.
Writing explicitely the lowest  order 
terms   one finds: 
\begin{equation}
P_{\beta \to \alpha} =  [{\hat h}_{\alpha \beta} \, 
{\hat h}^*_{\alpha \beta}]~y^2 
+ {\rm Im}[{\hat h}^*_{\alpha \beta} \; ({\hat h}^2)_{\alpha \beta}]~y^3 +
\ldots
\label{eq:expand}
\end{equation}
from where we can read the expressions  for $c_2^{\alpha \to \beta}$ 
and $c_3^{\alpha \to \beta}$.
The symmetry  properties of the coefficients 
can be easily checked:
\begin{enumerate}
\item The    lowest  order  term ($\propto  y^2$) of the probability  
is symmetric for   time  reversal, since ${\cal H}_0$ is  an  Hermitean
matrix: $({\cal H}_0)_{\alpha \beta} = ({\cal H}_0)^*_{\beta \alpha}$.

\item The  leading term is also symmetric  for
a $CP$   transformation   since in this  case one  has to replace
the Hamiltonian ${\cal H}_0$ with    its  complex  conjugate.

\item  The next to  leading term  $\propto y^3$  is
antisymmetric under a  $CP$ or $T$  transformation
as can  be immediately deduced  from the 
fact that   ${\cal H}_0$ is  hermitean.
\end{enumerate}
The coefficients  $|{\hat h}_{\alpha \beta}|^2$  and
${\rm Im} [{\hat h}^*_{\alpha \beta}\, ({\hat h}^2)_{\alpha \beta}]$
can be  easily  written   in terms  of the mixing  parameters
and  the squared mass ratio  $x_{12}$
verifying  that  the expansion  (\ref{eq:expand})  is  identical  to the one 
given in equation (\ref{eq:vacuum-vh}).
 
\subsection{Oscillations in matter}
The effective  Hamiltonians  describing the   flavor  evolutions
of $\nu$'s and $\overline{\nu}$'s  in matter  can be  written as:
\begin{eqnarray}
{\cal H}_{\nu} \,L & = &  {\cal H}_{0} \,L + {\hat p}_e \,V\, L =  
{\hat h} \, y + {\hat p}_e \, z 
\\
{\cal H}_{\overline{\nu}} \,L & = &  {\cal H}_{0}^* \,L - {\hat p}_e \, V\, L =  
{\hat h}^* \, y - {\hat p}_e \, z 
\end{eqnarray}
where  we  have introduced  the $\nu_e$  projection operator
${\hat p}_e$  (with the  property (${\hat p}_e)^n = {\hat p}_e$)
 that in the flavor  basis  has the components
$({\hat p}_e)_{\alpha \beta} = \delta_{\alpha e} \delta_{\beta e}$ and
a second adimensional  quantity:
\begin{equation}
z = V\, L
\end{equation}

Writing the  $S$ matrix   in  an  expanded  form, squaring 
the element $S_{\beta \alpha}$  and collecting  all terms
proportional to  $(y^n\, z^m)$  one can obtain  the 
transition probability as  a  power  series in  both  $y$ and $z$:
\begin{equation}
P_{\alpha \to \beta} (L, E_\nu) = 
P_{\alpha \to \beta} (y, z) = \sum_{n=2}^\infty \,\sum_{m=0}^\infty \,
c_{n, m}^{\alpha \to \beta} ~y^n\, z^m
\label{eq:power-matter}
\end{equation}
Naively one could  expect to  see terms  of  order
$(y^0 \, z^2)$,   $(y\, z)$  and $(y \, z^2)$   but they are  present 
only on the diagonal of the $S$ matrix  and are irrelevant for the transition
probabilities.

Note that  the set  of coefficients 
$c^{\alpha \to \beta}_{n, 0}$    are of course identical to 
the vacuum expansion.
From this we can deduce  the  very important  fact: 
if $z$ is  small,   that is  when the  $\nu$  pathlength is
much  shorter  than the  matter length 
$V^{-1}$,  the oscillation probabilities  are
approximately equal  to  the vacuum  case.
This  is  a sense is  not an entirely obvious  fact, because
a condition on the pathlength   does not  say anything  about the
importance  of the  matter effects on the neutrino masses and
mixing. It is  indeed  possible that  $E_\nu$  and  $V$ are such  that the 
mixing parameters  are  entirely  different  from the vacuum case 
(for  example   one  could sit on a MSW resonance),  however, if 
$L \ll V^{-1}$, the   oscillation probabilities   will
coincide with the vacuum  case.
This result,  especially  in the 3$\nu$ case,
appear as the   consequence of some  remarkable 
``cancellations''   between the  effective  values of the mixing 
parameters and squared masses,   however in this  formalism it is
entirely natural and  obvious.

Some important  symmetry  of the coefficients are the following:
\begin{equation}
 \cases { 
c_{n,m}^{\alpha \to \beta} =
+c_{n,m}^{\beta \to \alpha} =
+c_{n,m}^{\overline {\alpha} \to \overline {\beta}} =
+c_{n,m}^{\overline {\beta} \to \overline {\alpha}}
~~~~~          &for $n$~even and $m$ even, \cr
\omit      & \omit \cr
c_{n,m}^{\alpha \to \beta} =
+c_{n,m}^{\beta \to \alpha} =
-c_{n,m}^{\overline {\alpha} \to \overline {\beta}} =
-c_{n,m}^{\overline {\beta} \to \overline {\alpha}}
~~~~~          &for $n$~even and $m$ odd \cr
\omit      & \omit \cr
c_{n,m}^{\alpha \to \beta} =
-c_{n,m}^{\beta \to \alpha} =
-c_{n,m}^{\overline {\alpha} \to \overline {\beta}} =
+c_{n,m}^{\overline {\beta} \to \overline {\alpha}}
~~~~~          &for $n$~odd and $m$ even \cr
\omit      & \omit \cr
c_{n,m}^{\alpha \to \beta} =
-c_{n,m}^{\beta \to \alpha} =
+c_{n,m}^{\overline {\alpha} \to \overline {\beta}} =
-c_{n,m}^{\overline {\beta} \to \overline {\alpha}}
~~~~~          &for $n$~odd and $m$ odd \cr
 }
\end{equation}
Note  how the matter  effects  when they enter with an 
odd  power  of the  potential    have opposite  signs 
for $\nu$ and $\overline{\nu}$.

An  explicit  calculation of the   coefficients of lower order 
in $y$  gives   for  $n=2$:
\begin{equation} 
c_{2,m({\rm odd})}^{\alpha \to \beta}  = 0
\end{equation}
while  for   $m$    even  one has:
\begin{equation}
 c_{2,m ({\rm even})}^{\alpha \to \beta}  = 
[{\hat h}_{\alpha \beta}\, {\hat t}^*_{\alpha \beta}]~
d_{2,m} 
\end{equation}
where   $d_{2, m}$ are simple  numerical  coefficients:
\begin{equation}
d_{2,m({\rm even})}  =
 i^m \; \sum_{k=0}^m {(-1)^k  \over (k+1)! \, (m+1 -k)!}
 = { 2 \over (m+2)!}~ i^m
\end{equation}
For the  elements  with $n=3$  one  has:
\begin{equation}
c^{\alpha \to \beta}_{3, m({\rm even})} =
{\rm Im}[{\hat h}^*_{\alpha \beta} \; ({\hat h}^2)_{\alpha \beta}]
~d_{3,m}
\end{equation}
\begin{equation}
c^{\alpha \to \beta}_{3, m({\rm odd})} =
{1 \over 6} \, {\rm Re} \{ {\hat h}^*_{\alpha \beta} \;
[2\, {\hat h}_{\alpha e} \, {\hat h}_{e \beta} - ({\hat h}^2)_{\alpha \beta}
\,(\delta_{\alpha e} + \delta_{\beta e}) ] \}
~d_{3,m}  
\end{equation}
where  the quantities  $d_{3, m}$ are numerical  coefficients:
\begin{equation}
 d_{3,m{(\rm even})}  = 
2\,i^m ~\sum_{k=0}^m {(-1)^k  \over (k+1)! \, (m+2 -k)!}
 = { 2 \over (m+2)!}~ i^m 
 ~~~~  {\rm for}~  m~ {\rm even} 
\end{equation}
\begin{equation}
 d_{3,m}  = 
12 \, i^{m-1} ~\sum_{k=0}^m {(-1)^k  \over (k+1)! \, (m+2 -k)!}
 =  { 12(m+1) \over (m+3)!}~ i^{m-1}
~~~~  {\rm for}~  m ~{\rm odd}
\end{equation}
(The coefficients  have been defined so that
$d_{2,0} = d_{3,0} = d_{3,1} = 1$).

It is  useful, for example  when considering  an experiment with a fixed
baseline, to  resum over all $z$  terms, and  express the probability
again as  a  power series in $y$  (that  corresponds then to a power series
in $E_\nu^{-1}$), with coefficients  that are distance  dependent:
For the lowest  terms the  sum is easily obtained:
\begin{equation}
\sum_{m=0}^\infty  d_{2,m} \, z^m  =
\sum_{m({\rm even})=0}^\infty  d_{2,m} \, z^m =
\left ( {2 \over z} \right )^2 
\sin^2 \left ({z \over 2} \right ) 
\end{equation}
\begin{equation}
\sum_{m({\rm odd})=1}^\infty  d_{2,m} \, z^m  = 
{48 \over z^3 } ~
\left [  
\sin^2 \left ({z \over 2} \right ) - {z \over 4} \sin (z) \right ] 
\end{equation}
The reason  to keep separate  the sums   of
 the   $m$--even and $m$-odd  contributions,
  is   because they have different  symmetry properties
under a $CP$   transformation.
 For  $m$ even (odd)   the  
 $c_{n,m}^{\alpha \to \beta}$ coefficients   have the same
(opposite)  sign for  $\nu$'s  and  $\overline{\nu}$'s.

It can be interesting to verify   the results 
derived above in the special case
of two neutrino  mixing, when simple exact  expressions
for  oscillations probabilities exist.
This is  done in the next section.

\newpage
\section* {Appendix  B: Two flavor  mixing case}
\addtocounter{section}{1}

It can  be useful to    to discuss 
$\nu_e \leftrightarrow \nu_\mu$ or  $\nu_e \leftrightarrow \nu_\tau$ 
oscillations
in the  case of two flavor oscillations  when
the explicit calculation of the  oscillation probabilities
in matter is  very simple. 
Moreover this   case  is  an  exact solution in two  interesting
limiting cases:
\begin{enumerate}
\item The limit $x_{12} \to 0$  ($\Delta m^2_{12}$ small).
In this  case one has 
to replace  $\theta \to \theta_{13}$,
$\Delta m^2 \to \Delta m^2_{23}$.
The oscillation  probabilities for the 
$\nu_e \leftrightarrow \nu_\mu$  and 
$\nu_e \leftrightarrow \nu_\tau$  transitions are then  given by the
two flavor formula    multiplying by
$\sin^2 \theta_{23}$ and
$\cos^2 \theta_{23}$. 

\item  The limit  $\theta_{13} \to 0$.
In this case  one has to 
make the replacements
$\theta \to \theta_{12}$,
$\Delta m^2 \to \Delta m^2_{12}$.
The oscillation  probabilities for the 
$\nu_e \leftrightarrow \nu_\mu$  and 
$\nu_e \leftrightarrow \nu_\tau$  transitions are then  given by the
two flavor formula    multiplying by
$\cos^2 \theta_{23}$ and
$\sin^2 \theta_{23}$. 
\end{enumerate}

In the two flavor  approximation the vacuum
 oscillation probability  is:
\begin{equation}
P_{\rm vac} (\nu_e \to \nu_\mu) = 
\sin^2 2 \theta 
\sin^2 \left [ {\Delta m^2 \,L \over 4 \,E_\nu} \right ]
\label{eq:p2v}
\end{equation}
To have the oscillations in matter we simply have 
perform the  replacements
$\theta \to  \theta_{\rm m}$ and $\Delta m^2 \to (\Delta
m^2)_{\rm m}$, 
with: 
\begin{equation}
\sin^2 2 \theta_{\rm m}  = 
\sin^2 2 \theta ~ \left [\sin^2 2 \theta 
+ \left ( \cos 2 \theta \mp  {2 E_\nu \, V \over \Delta m_{23}^2} 
 \right )^2  \right ]^{-1}
\label{eq:p2s}
\end{equation}
and
\begin{equation}
(\Delta m^2)_{\rm m} = \Delta m^2 ~
\sqrt {\sin^2 2 \theta + \left ( \cos 2 \theta \mp 
{2 E_\nu \, V \over \Delta m^2}  \right )^2 }
\label{eq:p2d}
\end{equation}
The  minus  (plus)  sign  refers to neutrinos  (anti--neutrinos).
It can be seen that in  general:
(i) the oscillation probabilities in vacuum and in matter
can be  very  different from each other;
(ii) the oscillation probabilities  in matter  for 
$\nu$ and $\overline{\nu}$'s  are  also in general very different.

We are interested in the probability  for 
large   $E_\nu$. 
For   vacuum   oscillations  it is  straightforward,
to  expand  the probability as  a power series in 
$y = \Delta m^2  L /(4 E_\nu)$: 
\begin{equation}
P_{\rm vac} \simeq  \sin^2 2 \theta  
~\left ( {\Delta m^2 \, L \over  4 E_\nu } \right )^2  
- {\sin^2 2 \theta \over 3}
~\left ( {\Delta m^2 \, L \over  4 E_\nu } \right )^4  
+ \ldots
\end{equation}
To   compute  the same expansion for the  probabilities in matter
it is useful to
introduce the   variable:
\begin{equation}
\varepsilon = {\Delta m^2 \over 2 V E_\nu},
\end{equation}
using the shortened  notation 
$s = \sin 2 \theta$, $c = \cos 2 \theta$, the oscillation
probability in matter can then be rewritten as:
\begin{equation}
P_{\rm mat} (\nu_e \to \nu_\mu)  = 
{s^2  ~\varepsilon^2 \over 
1 \mp 2  \, c \, \varepsilon + \varepsilon^2} ~
\sin^2 \left [ {V L\over 2} ~\sqrt{  1 \mp 2 \, c \, \varepsilon +
\varepsilon^2 }  \right ].
\end{equation}
Developing in  a power  series in $\varepsilon$
and  writing  for  simplicity  $\alpha = {VL \over 2}$ one  obtains:
\begin{equation}
P_{\rm mat} (\nu_e \to \nu_\mu) =
s^2 \;  \varepsilon^2 ~ \sin^2 \alpha
 \pm  2\, \; {s^2 c} \;  \varepsilon^3
 ~ (\sin^2 \alpha - \alpha
 ~\sin\alpha \cos\alpha)  + O(\varepsilon^4)
\label{eq:prob2a}
\end{equation}
that  can be rewritten as:
\begin{eqnarray}
P_{\rm mat} (\nu_e \to \nu_\mu)
& = & \sin^2 2 \theta  
~\left ( {\Delta m^2 \, L \over  4 E_\nu } \right )^2  
~\left [ {1 \over \alpha^2}  ~\sin^2 \alpha \right ]
   \nonumber \\
& +  & {\sin^2 2 \theta  \, \cos 2 \theta \over 3} ~
~\left ( {\Delta m^2 \, L \over  4 E_\nu } \right )^3  
~\left [ \left ({6 \over \alpha^3} \right )
( \sin^2  \alpha - \alpha \sin\alpha \cos \alpha) \right  ]
\label{eq:prob2b} \\
 & + &  \ldots \nonumber  
\end{eqnarray}
We can note that the first  (second) term   is  symmetric
(anti--symmetric) for the replacement $V \to -V$
that is  replacing 
 $\nu$ with  $\overline{\nu}$.

Developing the  expression (\ref{eq:prob2b})
 for   small $\alpha$ 
(that is  for   $L < V^{-1}$) and  reinserting  the definition
one obtains:
\begin{eqnarray}
P_{\rm mat} (\nu_e \to \nu_\mu)
& = & \sin^2 2 \theta  ~\left ( {\Delta m^2 \, L \over  4 E_\nu } \right )^2  
\left [ 1  -   {(VL)^2 ~ \over 12}  + \ldots  \right ]    \nonumber \\
& +  & {\sin^2 2 \theta  \, \cos 2 \theta \over 3} ~
 \left ( {\Delta m^2 \, L \over  4 E_\nu } \right )^3  ~VL~
\left [ 1  -   {(VL)^2 ~ \over 15}  + \ldots  \right ]   
\\ & + &  \ldots \nonumber  
\label{eq:prob2c}
\end{eqnarray}
We have  obtained  with an explicit calculation
a set  of interesting results  for the
oscillation probability in the limit of large  energy
\begin{itemize}
\item [(i)] The oscillation probabilities for  $\nu$'s  and  $\overline{\nu}$'s
  become asymptotically   equal,  and    vanish with a
  leading contribution of order $y^2 \sim E_\nu^{-2}$.

\item [(ii)]  The matter effects  generate  a correction
of opposite sign for $\nu$'s  and  $\overline{\nu}$'s.  This  correction
vanishes  more rapidly with   increasing energy $\propto  y^3 \sim E_\nu^{-3}$

\item [(iii)] 
For  small  pathlength:  $L \aprle V^{-1}$
the matter  effects  grow  linearly with   the matter  potential  $V$.

\item [(iv)] Again for  small  pathlength 
the leading order   term  of the oscillation probability   grows
with the pathlength  $\propto L^{2}$, while the correction
due to the matter  effects
grows  more rapidly  $\propto L^{4}$.
\end{itemize}

Comparing equations (\ref{eq:prob2b}) and (\ref{eq:prob2c})
with equations (\ref{eq:prob-asymptotic})
and (\ref{eq:prob-asymptotic1})
one can see  that we have  reproduced 
with an  explicit calculation  the results   
for the leading  order term of the oscillation probability
and for  the  matter  effects. 
It is also  easy to see  that  in the  two  flavor  case considered
here the  free Hamiltonian, and its  square 
  can be written  in the flavor basis as:
\begin{equation}
{\cal H}_0 = {\Delta m^2 \over 4 E_\nu} ~\left [
\begin{array}{cc}
 -\cos 2 \theta       &  \sin 2 \theta   \\
 \sin 2 \theta   &   \cos 2 \theta 
\end{array}
\right ], 
~~~~~~
({\cal H}_0)^2 = \left( {\Delta m^2 \over 4 E_\nu} \right )^2 ~\left [
\begin{array}{cc}
 1       &  0  \\
 0   &   1 
\end{array}
\right ]
\end{equation}
Calculating the coefficient $A_{e\mu}$, $B_{e \mu}$ and $C_{e \mu}$
according  to the general  formulae 
 (\ref{eq:A}),
 (\ref{eq:B})  and
 (\ref{eq:C})  one obtains:
$A = \sin^2 2 \theta$,  $B = 0$  
($CP$  and $T$ violation 
effects vanish,  in the  two flavor case)
and
$C = \cos 2 \theta\, \sin^2 2 \theta/3$  in agreement with the 
general result.

\newpage
\section* {Appendix C: 
Masses  and mixing   for  large  matter effects}
\addtocounter{section}{1}

\begin{figure} [t]
\centerline{\psfig{figure=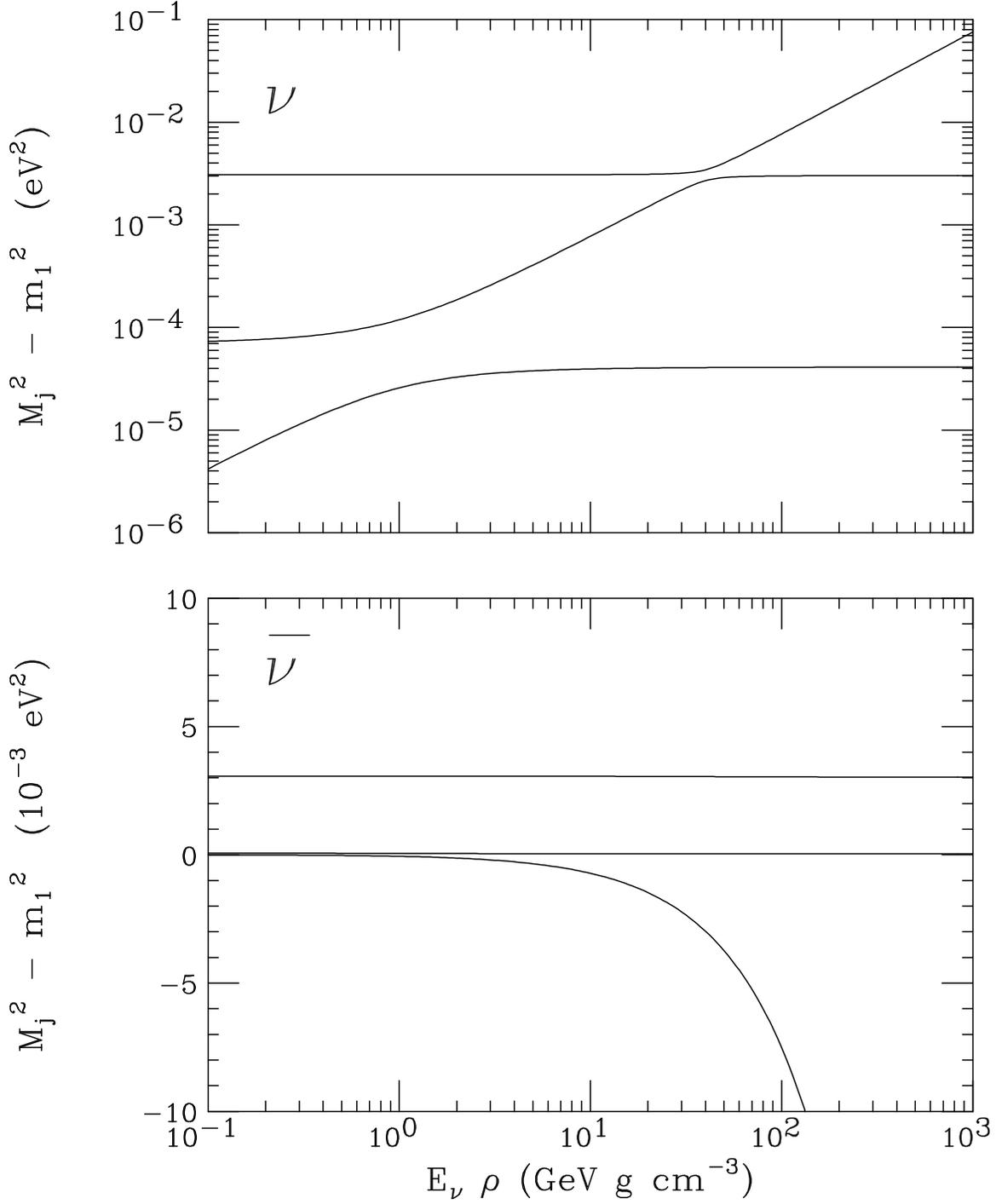,angle=90,height=19.0cm}}
 
\caption {Effective  squared mass eigenvalues in matter, plotted as
a  function of  $E_\nu \, \rho$,
The  top (bottom) panel is for $\nu$ ($\overline{\nu}$).
The   squared mass values are: $m_1^2 = 0$,
$m_2^2 = 7 \times 10^{-5}$~eV$^2$,
$m_3^2 = 3 \times 10^{-3}$~eV$^2$;
the mixing parameters are:
$\theta_{12} = 40^\circ$, 
$\theta_{23} = 45^\circ$, 
$\theta_{13} = 7^\circ$, 
$\delta = 45^\circ$.
In this  figure the elkec 
\label{fig:mass}  }
\end{figure}

\begin{figure} [t]
\centerline{\psfig{figure=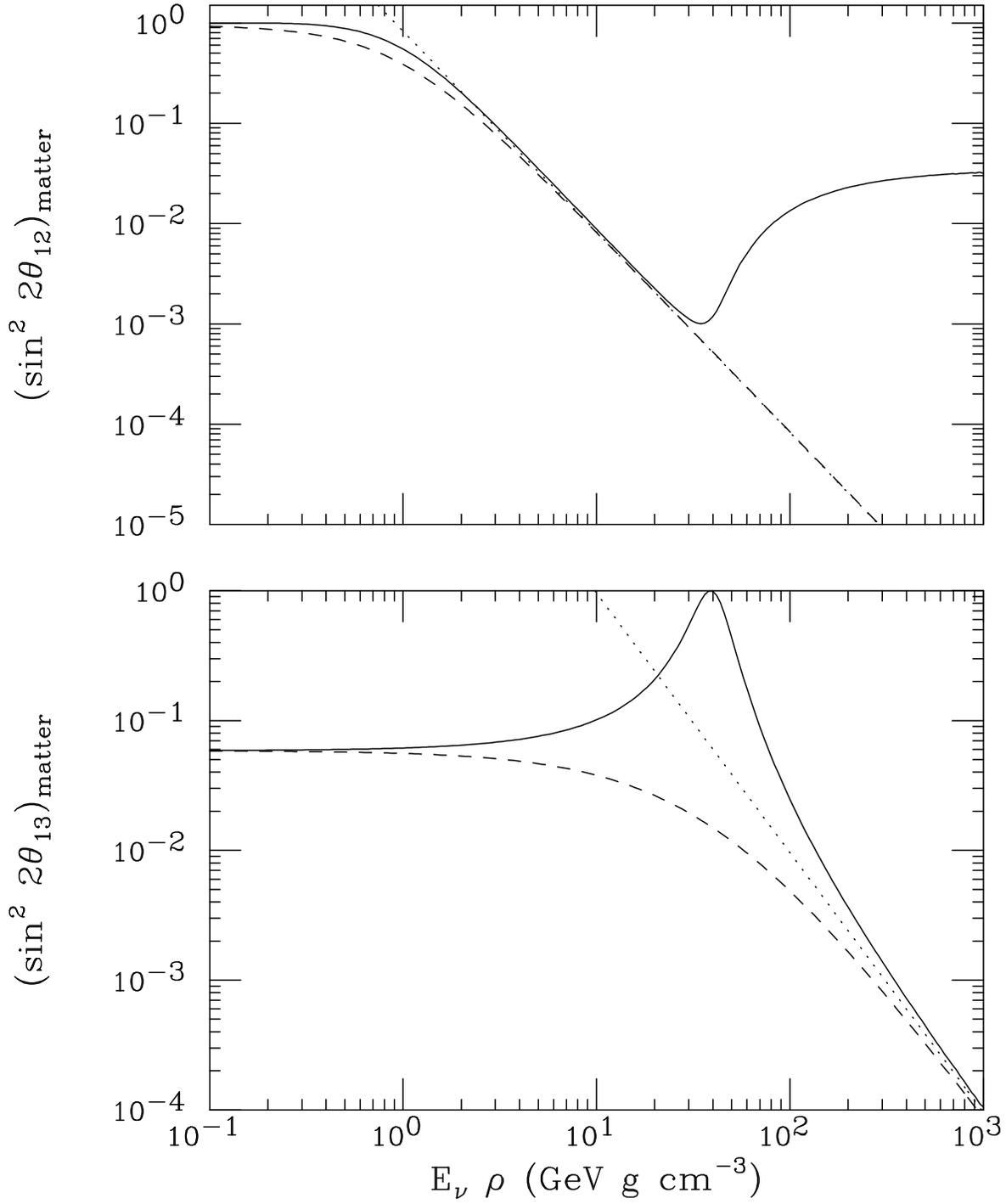,angle=90,height=19.0cm}}

\caption {Mixing parameters in matter as a  function  of the product
$E_\nu ~\rho$:  top panel $\sin^2 2 \theta_{12}^{\rm m}$,
bottom  panel
 $\sin^2 2 \theta_{13}^{\rm m}$.
The   squared mass values are: $m_1^2 = 0$,
$m_2^2 = 7 \times 10^{-5}$~eV$^2$,
$m_3^2 = 3 \times 10^{-3}$~eV$^2$;
the mixing parameters in vacuum are:
$\theta_{12} = 40^\circ$, 
$\theta_{23} = 45^\circ$, 
$\theta_{13} = 7^\circ$, 
$\delta = 45^\circ$.
The solid curves are for $\nu$'s, the dashed  curves for 
$\overline{\nu}$'s
Note the simple  asymptotic  forms  of the parameters for  large 
$E_\nu \, \rho$.
The  dotted  lines    in the top panel is 
$\sin^2 2 \theta_{12} \, (\Delta m^2_{12}/(2 E_\nu V))^2$.
The  dotted  line in the bottom  panel is 
$\sin^2 2 \theta_{13} \, (\Delta m^2_{13}/(2 E_\nu V))^2$.
\label{fig:mixing-parameters-1}  }
\end{figure}

\begin{figure} [t]
\centerline{\psfig{figure=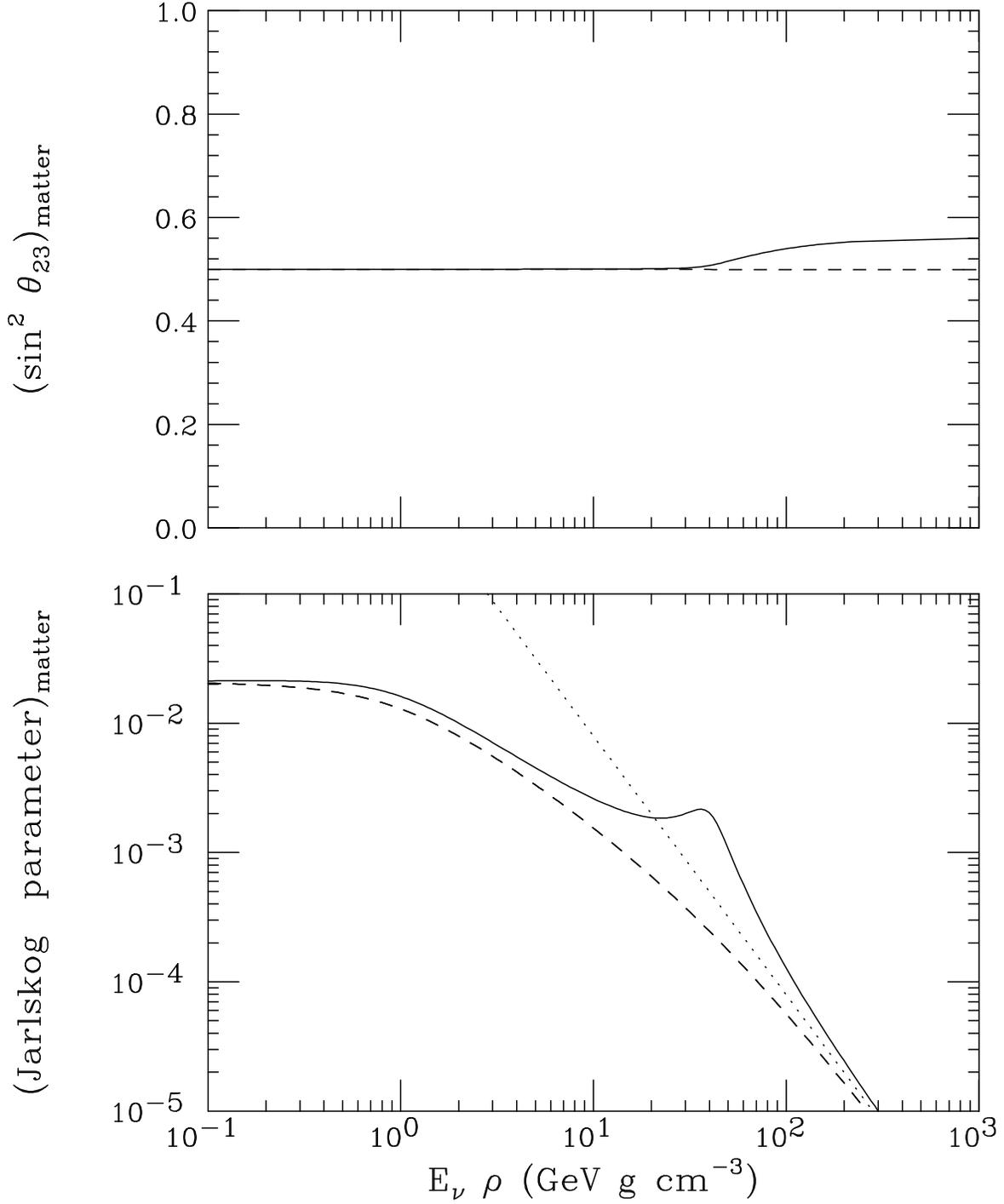,angle=90,height=19.0cm}}
 
\caption {Effective mixing parameters in matter plotted as a  function  
of the product
$E_\nu ~\rho$.  The top panel   shows $\sin^2 \theta_{23}^{\rm m}$,
 the bottom panel shows  the Jarlskog parameter:
$J = (c_{13}^{\rm m})^2 \, s_{13}^{\rm m} \, s_{12}^{\rm m} 
\, c_{12}^{\rm m} \,
 s_{23}^{\rm m} \, c_{23}^{\rm m} \, \sin \delta_{\rm m}$.
The   squared mass values are: $m_1^2 = 0$,
$m_2^2 = 7 \times 10^{-5}$~eV$^2$,
$m_3^2 = 3 \times 10^{-3}$~eV$^2$;
the mixing parameters in vacuum are:
$\theta_{12} = 40^\circ$, 
$\theta_{23} = 45^\circ$, 
$\theta_{13} = 7^\circ$, 
$\delta = 45^\circ$.
The solid curves are for $\nu$'s, the dashed  curves for 
$\overline{\nu}$'s
The  dotted  line in the  bottom  panel is 
$J_{\rm vac} \, \Delta m^2_{13} ~\Delta m^2_{12}/(2 E_\nu V)^2$
and  indicates the asymptotic  form  of  the  
parameter  for large  $E_\nu \, \rho$.  Note  how the asymptotic
form  is the same for $\nu$'s  and $\overline{\nu}$'s.
\label{fig:mixing-parameters-2}  }
\end{figure}

\begin{figure} [t]
\centerline{\psfig{figure=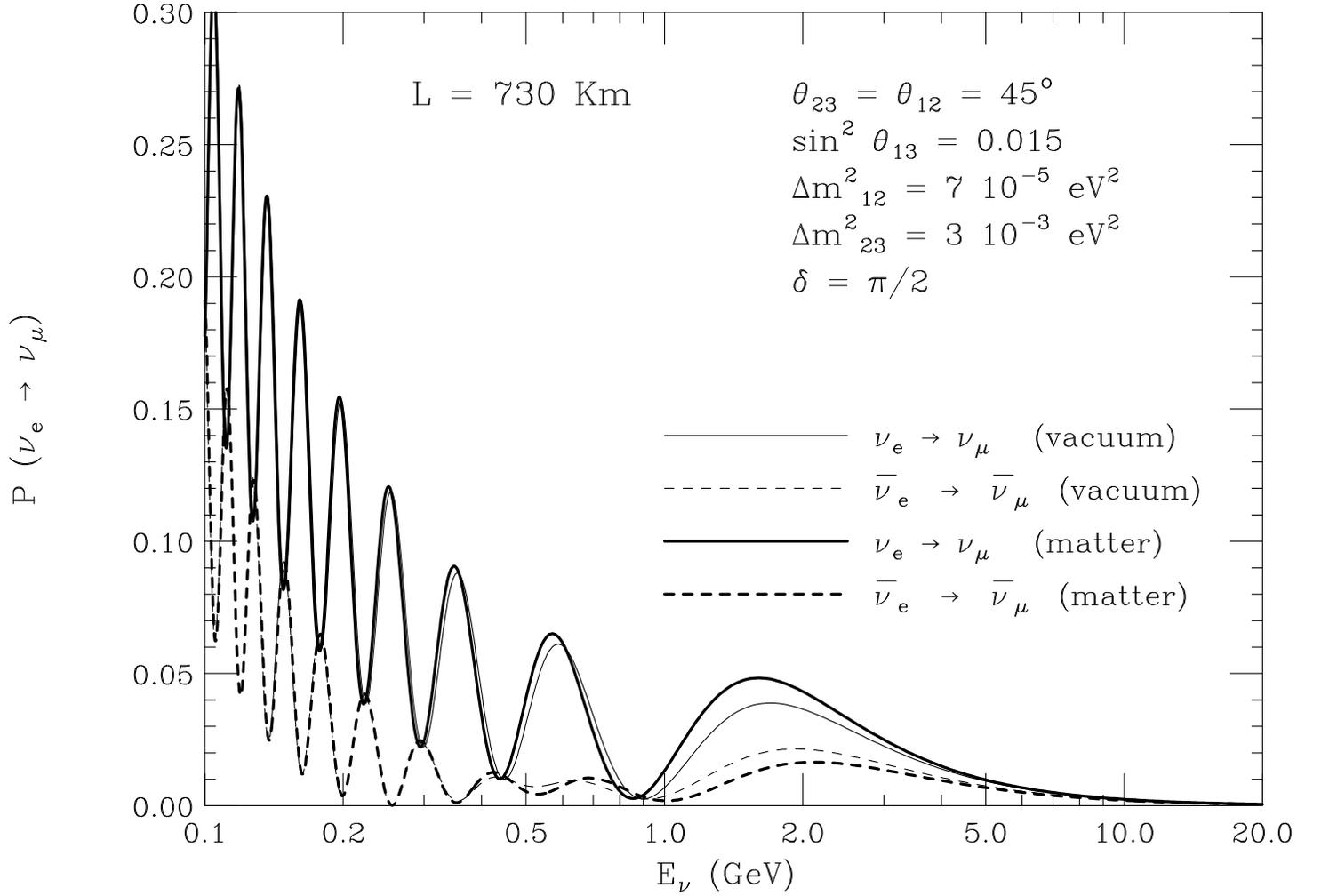,angle=90,height=13.5cm}}
 
\caption {Oscillation  probability   for the transition
$\nu_e \to \nu_\mu$  plotted as a function of $E_\nu$ for a fixed
value of the pathlength $L = 730$~Km.
The   oscillation parameters are  fixed,
and are given inside the figure. The four curves are for neutrinos 
and anti--neutrinos in vacuum and in matter.
\label{fig:prob_730}  }
\end{figure}

\begin{figure} [t]
\centerline{\psfig{figure=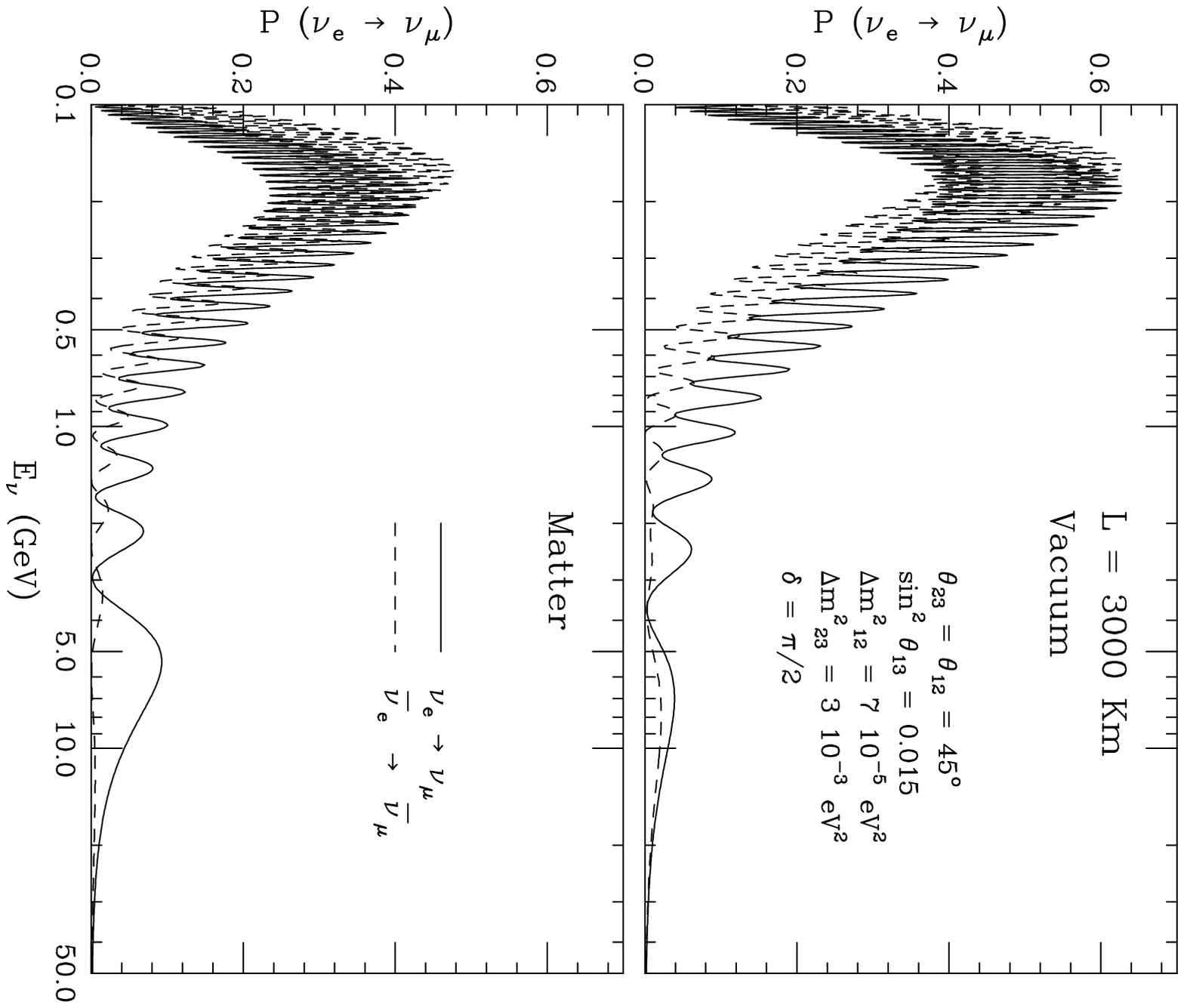,angle=90,height=19.0cm}}
 
\caption {Oscillation  probability   for the transition
$\nu_e \to \nu_\mu$  plotted as a function of $E_\nu$ for a fixed
value of the pathlength $L = 3000$~Km.
The   oscillation parameters are  fixed,
and are indicated  in the figure. The solid (dashed) curves are for $\nu$
($\overline{\nu}$). The top  (bottom)  panel  gives
the probability for propagation in vacuum (matter).
\label{fig:prob_3000}  }
\end{figure}

\begin{figure} [t]
\centerline{\psfig{figure=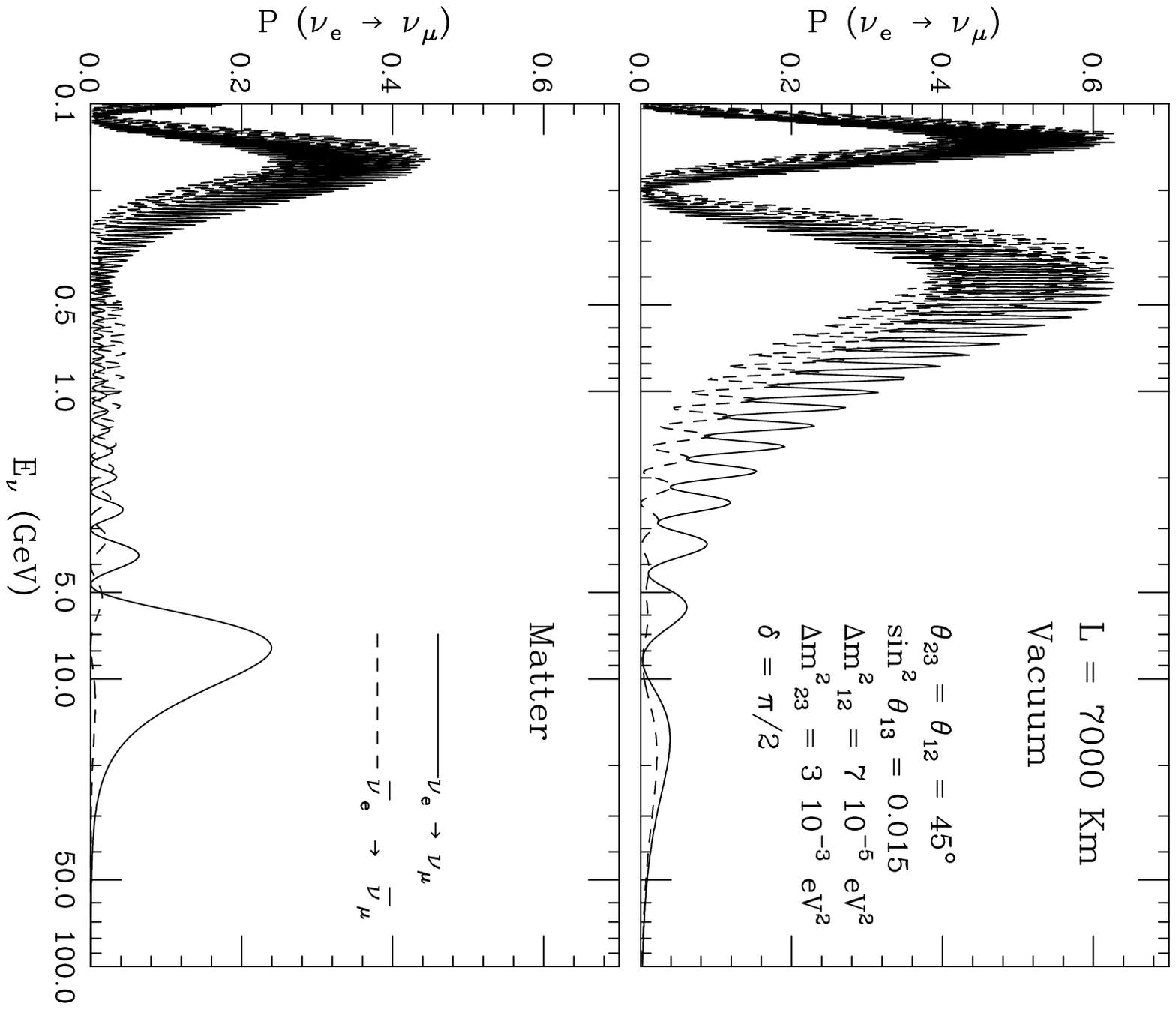,angle=90,height=19.0cm}}
\caption {Oscillation  probability   for the transition
$\nu_e \to \nu_\mu$  plotted as a function of $E_\nu$ for a fixed
value of the pathlength $L = 7000$~Km.
The   oscillation parameters are  fixed,
and are indicated  in the figure. The solid (dashed) curves are for $\nu$
($\overline{\nu}$). The top  (bottom)  panel  gives
the probability for propagation in vacuum (matter).
\label{fig:prob_7000}  }
\end{figure}

\begin{figure} [t]
\centerline{\psfig{figure=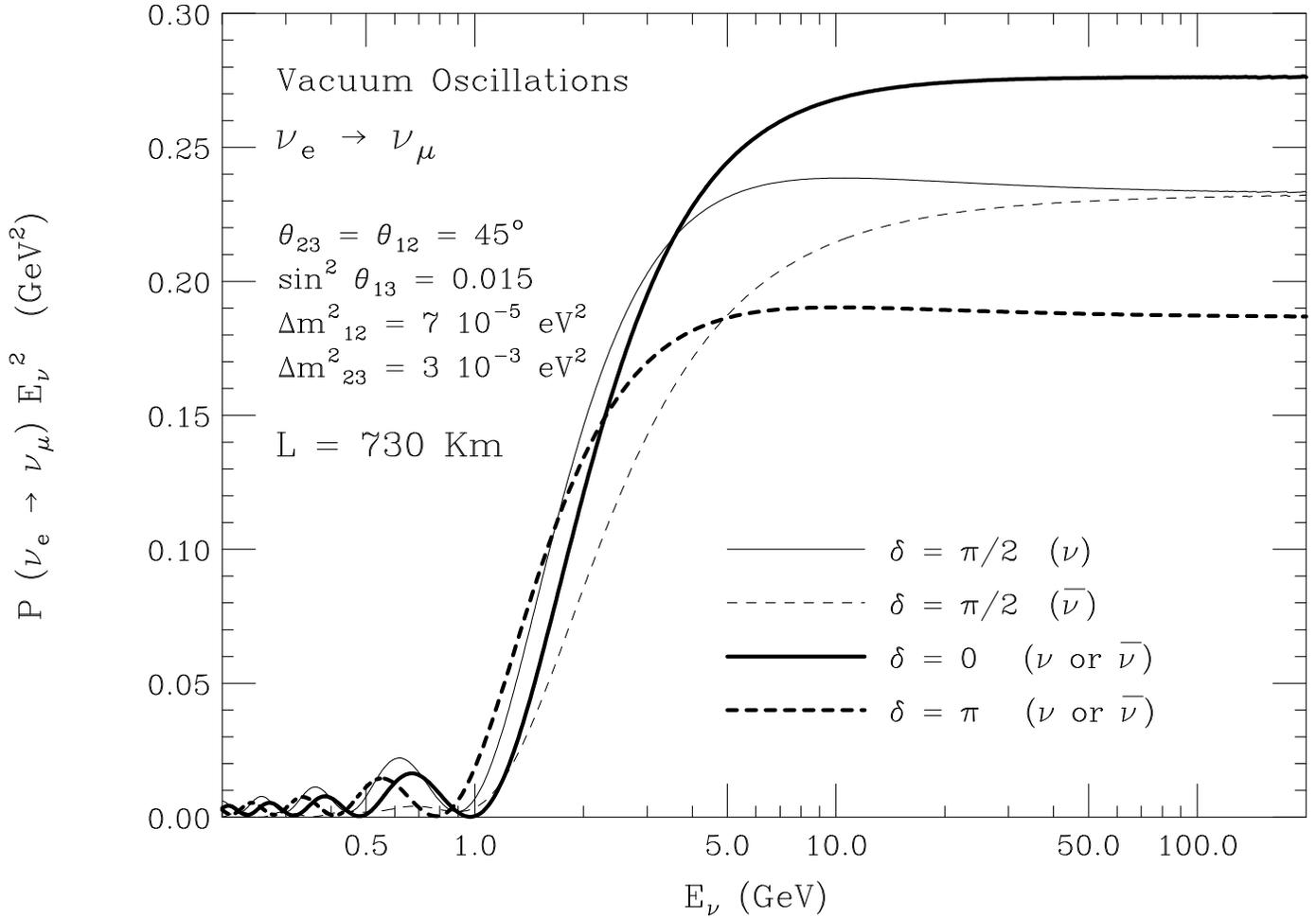,angle=90,height=13cm}}
 
\caption {In this  figures  we plot 
 as  a function of  $E_\nu$ 
(for a fixed value $L = 730$~Km) 
the product 
$P(\nu_e \to \nu_\mu) \times E_\nu^2$.
The different curves  correspond to  different  values of the
phase $\delta$.
The values of the other parameters  is indicated inside the  plot.
\label{fig:eprob0_730}  }
\end{figure}

\begin{figure} [t]
\centerline{\psfig{figure=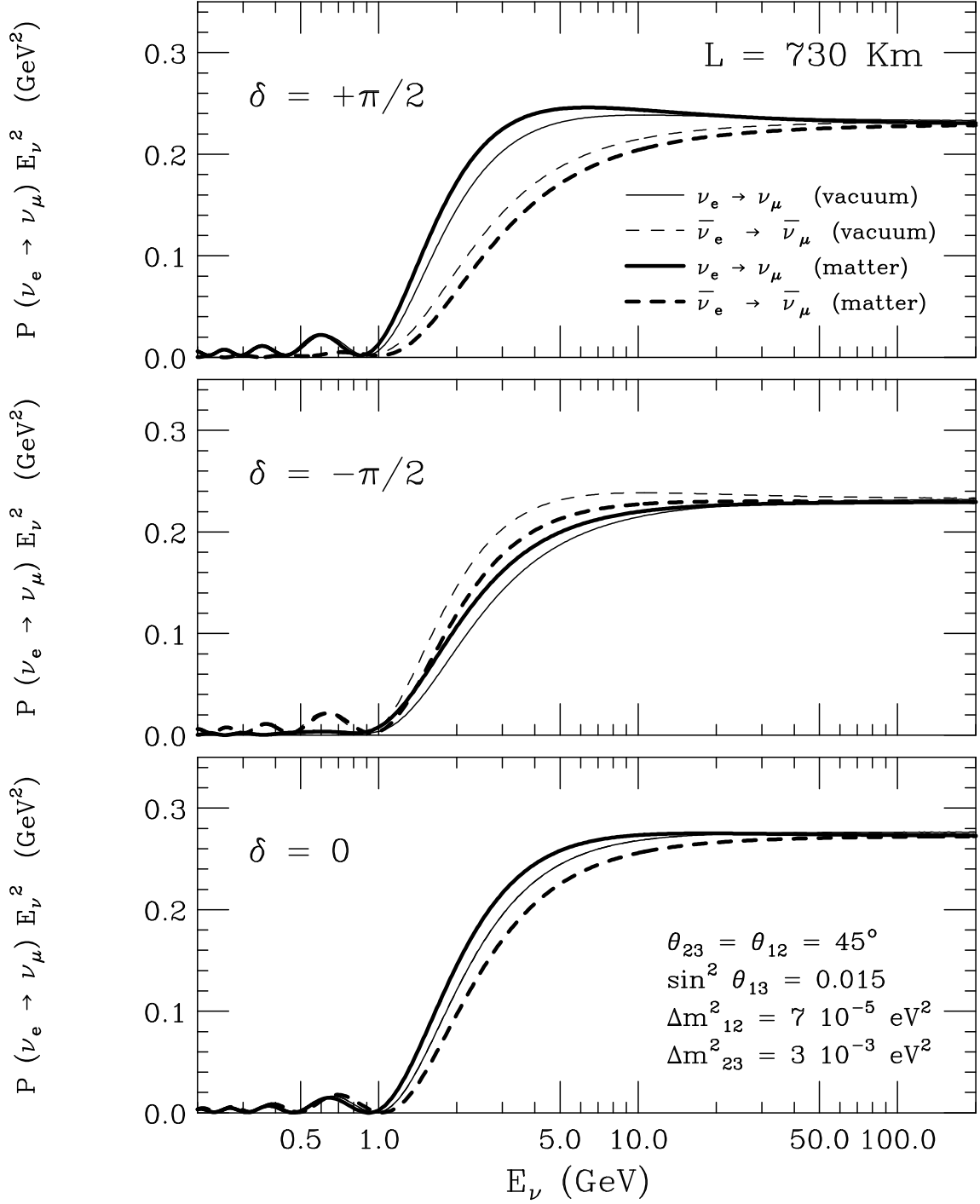,angle=90,height=19cm}}

\caption {In this  figure  we plot 
 as  a function of the neutrino
energy $E_\nu$ 
(for a fixed value $L = 730$~Km) 
the product 
$P(\nu_e \to \nu_\mu) \times E_\nu^2$.
The different curve  describe   the probability with
and  without matter  effects.
\label{fig:eprobm_730}  }
\end{figure}

\begin{figure} [t]
\centerline{\psfig{figure=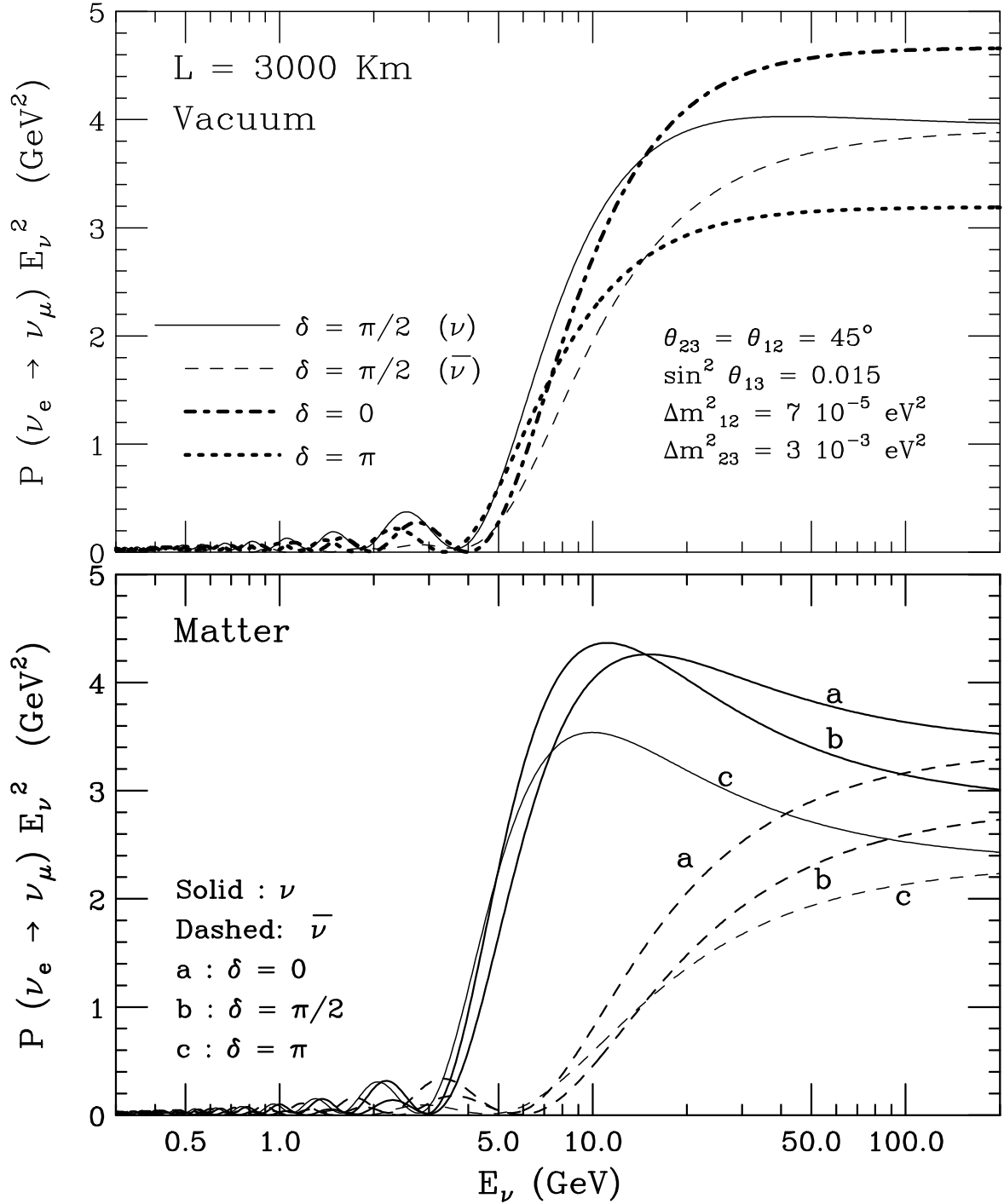,angle=90,height=19.0cm}}
 
\caption {In this  figures  we plot 
 as  a function of the neutrino
energy $E_\nu$ 
the product 
$P(\nu_e \to \nu_\mu) \times E_\nu^2$,
for a fixed value of the neutrino  pathlength $L = 3000$~Km.
The top panel is for  vacuum  oscillations
the bottom one includes  matter effects.
\label{fig:eprob_3000}  }
\end{figure}

\begin{figure} [t]
\centerline{\psfig{figure=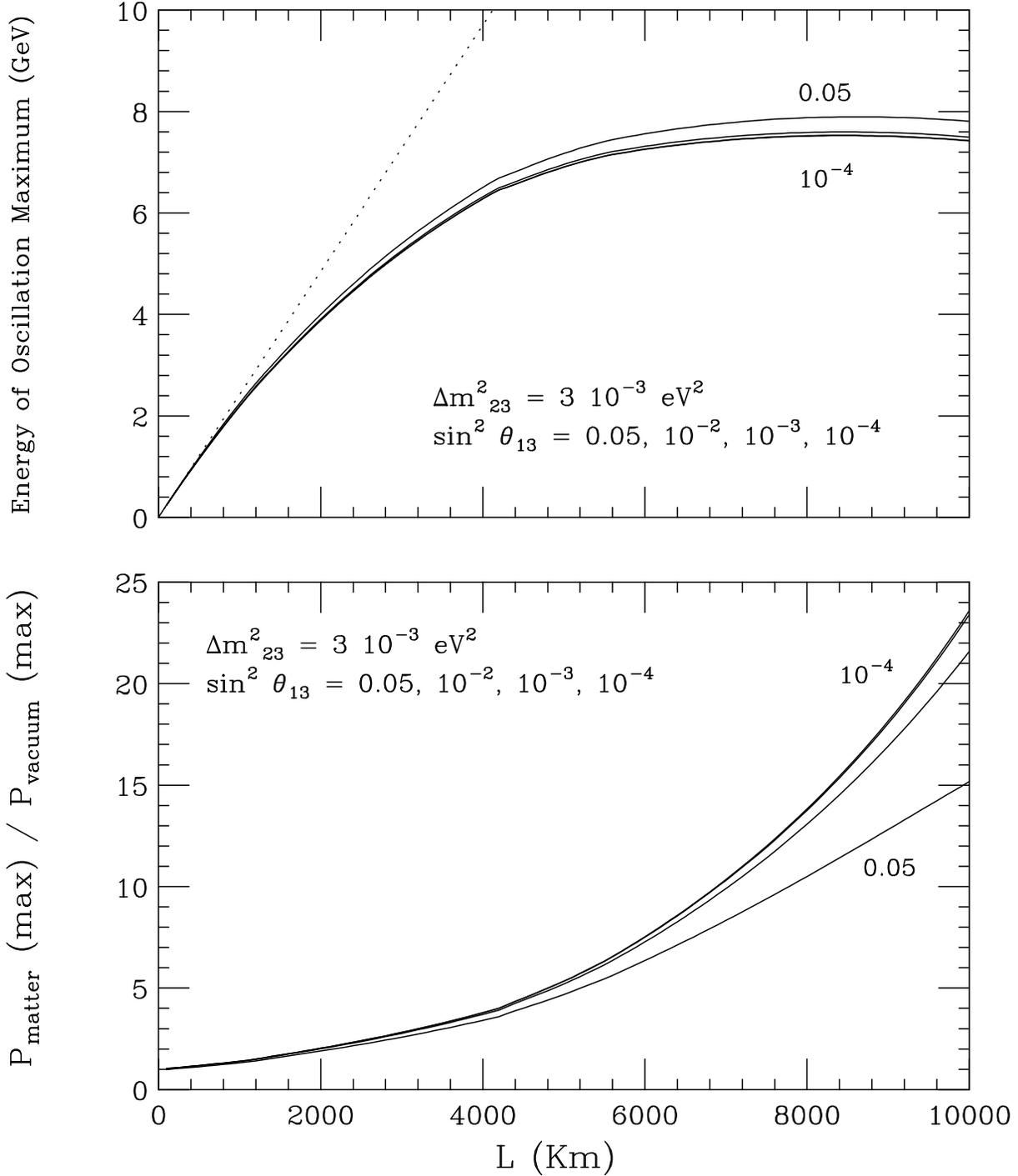,angle=90,height=19.0cm}}

\caption {The top  panel shows   as  a function of  the 
$\nu$ pathlength $L$,  the energy $E_{\rm  peak}$  where the 
 probabilities for
$\nu_e \leftrightarrow \nu_{\mu,\tau}$ transitions
have the   matter  enhanced  maximum.
The bottom  panel  show  the  size of the matter  enhancement.
The different  curves  correspond to   different values of   $s_{13}$.
For  small $s_{13}$  the position of the peak and
the size of the enhancement  are  independent  from $s_{13}$.
\label{fig:s13}  }
\end{figure}

\begin{figure} [t]
\centerline{\psfig{figure=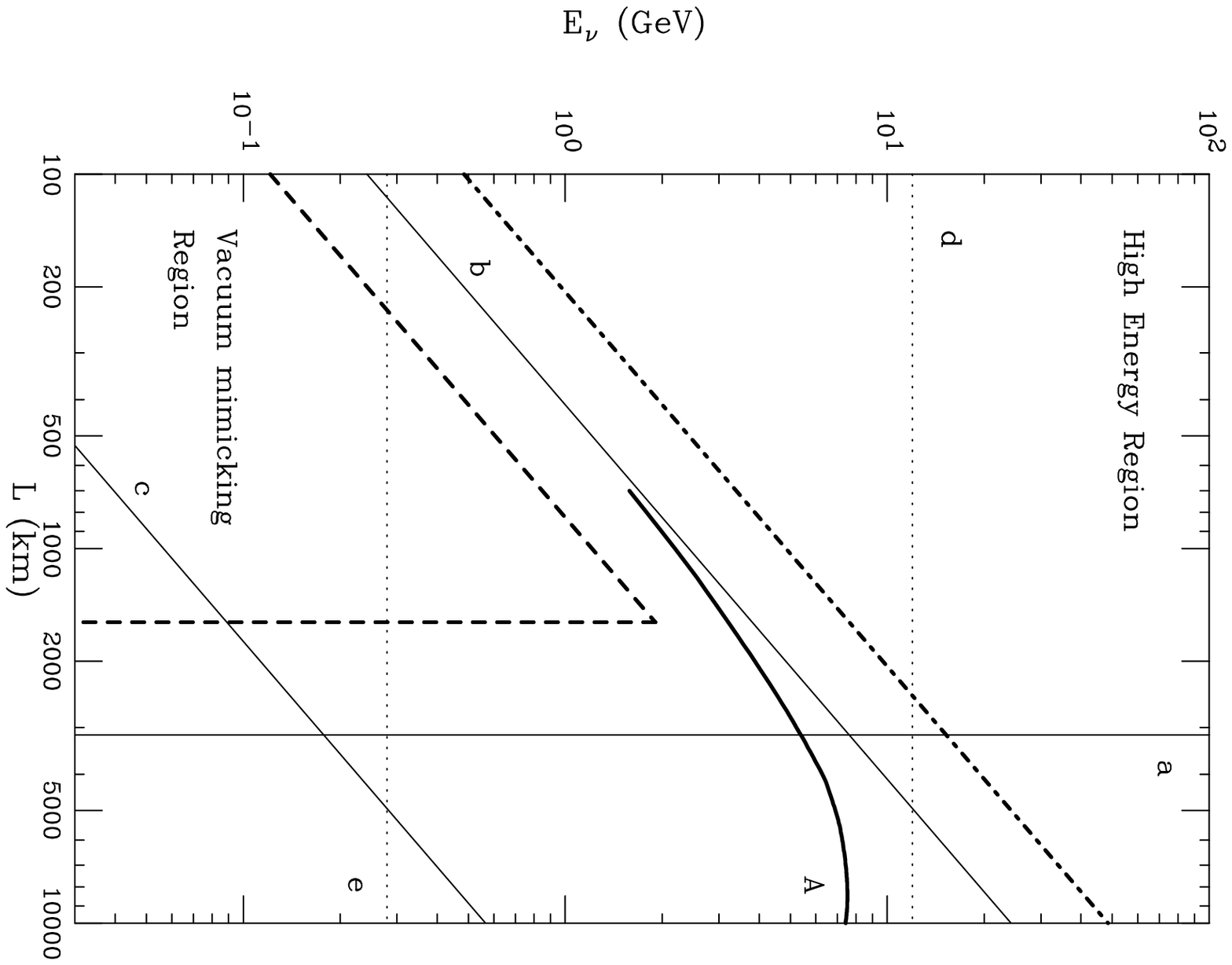,angle=90,height=17.cm}}
 
\caption {In this  figure are indicated some interesting
regions in the space  ($L$,$E_\nu$)  of the oscillation
probability for  $\nu_e \leftrightarrow \nu_\mu$
and  $\nu_e \leftrightarrow \nu_\tau$
transitions.
The region above the dot--dashed line
is  the one where  the  high energy expansion is  valid.
In the region delimited  by the dashed  line
the oscillation probabilities in matter and in vacuum
are to a good approximation equal.
The line  labeled  $A$  shows the $\nu$ energy where the 
oscillation probability has  the largest matter induced  enhancement.
The line  is drawn only if the maximum enhancement  is larger than 20\%.
The line  labeled with $a$   shows  the relation
$L = 2\, V^{-1}$  for   the Earth's crust  ($\rho = 2.8$~g~cm$^{-3}$).
The lines $b$ and $c$ show the relations
$E_\nu = |\Delta m_{23}^2| \, L/(2\pi)$ and
$E_\nu = \Delta m_{12}^2 \, L/(2\pi)$, that is the highest energy  
where  the  oscillation probability has a maximum.
Line  $d$  indicates the approximate energy for which 
$\theta_{13}$  passes through a resonance, and  line
$e$ shows  the energy above which matter  effects  modify
significantly the mixing parameters.
\label{fig:regions}  }
\end{figure}

The dependence
of the effective squared mass differences  and mixing parameters 
in matter  on  the product $\rho  E_\nu$  is shown
in figures~\ref{fig:mass}, \ref{fig:mixing-parameters-1}
and ~\ref{fig:mixing-parameters-2}
(in the figures we  assumed   an electron fraction 
of ${1 \over 2}$, therefore the density $\rho$ and the potential 
$V$ are  simply proportional).
Several important features are clearly visible:
\begin{enumerate}
\item The   squared mass eigenvalues and
of the mixing parameters  in matter are in general   very different for
 neutrinos and  anti-neutrinos.
Reversing  the sign of $\Delta m^2_{23}$, the   behaviour of $\nu$ and
$\overline{\nu}$'s to a good approximation is   simply exchanged.
\item All four  mixing  parameters   (the tree  angles and the phase)
change  when matter  is present,  however, when  the  two
mass scales  $\Delta  m^2_{12}$ and $|\Delta m^2_{23}|$ are
of different  orders of  magnitude, as  indicated by the 
the data,  the   angle 
$\theta_{23}$ and the phase $\delta$  remain approximately  constant
while the  angles  $\theta_{12}$ and $\theta_{13}$ change  
much more dramatically.
\item    We can recognize    several ranges of  the parameter
$a = 2 E_\nu V$  where the  behaviour of the solution  has 
different  characteristics:

\begin{itemize}
\item [(i)] When 
$2 \,E_\nu V \ll \Delta m^2_{12}$,
all matter  effects  are negligible
and the oscillations  develop as in the vacuum case.

\item [(ii)] For
$2 \,E_\nu V \ll |\Delta m^2_{23}|$
the  effective mass and 
 mixing  of  the state $\nu_3$   remain  unchanged, 
but
$\Delta m^2_{12}$ and  $\theta_{12}$ can be   modified by matter effects
(this is  interesting only if   $\Delta m^2_{12} \ll |\Delta m^2_{23}|$).
In this   situation it is a 
a good approximation to use the 
well known  two--flavor  formulas
to   obtain $(\Delta m^2_{12})^{\rm m}$ and  $\theta_{12}^{\rm m}$ 
as a function of  $2 V E_\nu$.

\item [(iv)] When  $2 \,E_\nu V  \simeq  |\Delta m^2_{23}|$
there is a resonance
($\theta_{13}^{\rm m}$ becomes  ${\pi \over 4}$).
The  resonance is present  for neutrinos  
if $\Delta m^2_{23}$ positive
or for  anti--neutrinos  if
$\Delta m^2_{23}$ is  negative.

\item [(v)] When 
$2 V E_\mu \gg |\Delta m^2_{23}|$, 
the behaviour of the mixing angles
and  squared  mass eigenvalues takes a simple form that 
 will be discussed  below.
\end{itemize}
\end{enumerate}

In  the study of the effective  parameters for large  matter effects
($|\Delta m^2_{23}| / (2 VE_\nu) \to 0$) 
one has  to distinguish two  cases:
\begin{itemize}

\item Case A corresponds  to neutrinos   for
$\Delta  m^2_{23}$ positive    or to anti--neutrinos
for $\Delta  m^2_{23}$  negative.
In this case one has: 
$$(\Delta m^2_{12})^{\rm m} \to  \Delta m^2_{23},$$
$$(\Delta m^2_{23})^{\rm m} \to 2 V E_\nu,$$
$$\sin^2\theta_{13}^{\rm m} \to 1, ~~~~
\cos^2 \theta_{13}^{\rm m} \to \sin^2 \theta_{13} \; \left (
{\Delta m^2_{23} \over 2
E_\nu V } \right )^2, $$
$$\sin^2 \theta_{12}^{\rm m} \to \sim 1,
~~~~~\cos^2 \theta_{12}^{\rm m} \to {\rm const} \simeq 
\sin^2 \theta_{12} ~{\Delta m^2_{12} \over \Delta m^2_{23}}.$$

Thin means that  $|\nu_3^{\rm m}\rangle$ 
 becomes asymptotically a  massive pure  $|\nu_e\rangle$ state, therefore
 $\theta_{13} \to  {\pi \over 2}$   and
$\cos^2 \theta_{13}$ vanishes  rapidly   $(\propto 2 E_\nu V)^{-2}$.
The  angle $\theta_{12}$  tends to
a  constant value,  reflecting the  fact that
the   $\nu_e$  flavor  is ``sucked away'' at the same  rate from the
$|\nu_1\rangle $ and $|\nu_2\rangle$ states.
The  squared mass difference
$(\Delta m^2_{23})^{\rm m}$   grows  without  bound
($ \to  2 E_\nu V$), 
while 
$(\Delta m^2_{12})^{\rm m}$    approaches the  constant value
 ($\to \Delta m^2_{23}$).

\item Case B corresponds  to neutrinos   for
$\Delta  m^2_{23}$   negative     or to anti--neutrinos
for $\Delta  m^2_{23}$  positive.
In this case one has: 
$$(\Delta m^2_{12})^{\rm m} \to 2 V E_\nu,$$ 
$$(\Delta m^2_{23})^{\rm m} \to  \Delta m^2_{23},$$
$$\sin^2\theta_{13}^{\rm m} \to \sin^2 \theta_{13} \; \left (
{\Delta m^2_{23} \over 2
E_\nu V} \right )^2, ~~~~~ \cos^2 \theta_{13}^{\rm m} \to 1,$$
$$\sin^2 \theta_{12}^{\rm m} \to 
\sin^2 \theta_{12} \left ( {\Delta m^2_{12} \over 2E_\nu V} 
\right )^2, ~~~~~ \cos^2 \theta_{12}^{\rm m} \to 1.$$

\end{itemize}
In this case 
it is   $|\nu_1^{\rm m}\rangle$ that   becomes  the massive
pue $|\nu_e\rangle$ state,
therefore   both   $\theta_{12}$ and   $\theta_{13}$ asymptotically
vanish  $\propto (2 E_\nu V)^{-1}$,    and 
it is $(\Delta m^2_{12})^{\rm m}$ that  grows  large
($ \to 2 E_\nu V$)  
while  $\Delta m^2_{23}$ remains  approximately  unchanged.
In both cases the modifications  to the 
the angle $\theta_{23}$ and the phase $\delta$   are small
and vanish for
$\Delta m^2_{12}/|\Delta m^2_{23}|$ small.

Note that the  behaviour of the  effective masses
and  mixing parameters 
is strikingly different    for    $\nu$'s  and $\overline{\nu}$'s.
However there are some important cancellations.
For example collecting the expressions  for the
different factors one can verify that the
the effective  Jarlskog parameter in matter 
($J_{\rm m} = 
(c_{13}^{\rm m})^2 \, s_{13}^{\rm m} \, s_{12}^{\rm m} \, c_{12}^{\rm m} \,
 s_{23}^{\rm m} \, c_{23}^{\rm m} \, \sin \delta_{\rm m}$)
is equal for $\nu$ and $\overline{\nu}$:
\begin{equation}
J_{\rm m}^{\nu} =
J_{\rm m}^{\overline{\nu}} \to  J_{\rm vacuum} ~
~ { \Delta m^2_{13}
 ~\Delta m^2_{12} \over (2 E_\nu V)^2},
\end{equation}
(see  for numerical calculation in  fig.~\ref{fig:mixing-parameters-2})
as  it is the case for 
 the product of the three effective squared mass differences:
\begin{equation}
(\Delta m^2_{12})_{\rm m}^{\nu} ~(\Delta m^2_{23})_{\rm m}^{\nu}
~(\Delta m^2_{13})_{\rm m}^{\nu}
=
(\Delta m^2_{12})_{\rm m}^{\overline{\nu}} ~(\Delta m^2_{23})_{\rm m}^{\overline{\nu}}
~(\Delta m^2_{13})_{\rm m}^{\overline{\nu}}
\to  \Delta m^2_{23} ~ (2 V E_\nu)^2
\end{equation}
Combining the  two  results  one satisfies  equation (\ref{eq:cancel}).

\newpage


\newpage

\newpage

\newpage

\newpage

\newpage

\newpage

\newpage

\newpage

\newpage

\newpage

%
%

\newpage

\newpage

\newpage

\newpage

\begin{thebibliography}{100}
\bibitem{tmu}
It should be noted
that   the study of $T$ violation, because of the
structure of the Earth's crust  could  require
two experiments with 
two accelerators (located  in points A and B) and two  detectors,
(located in  points B and A) to insure that also the 
matter  potential $V(t)$   that the neutrino  encounter during their
path  is replaced  by the $T$--transformed one.
This point  has  been   made  by H.~Nunokawa  and O.Yasuda
(private  communication).
\bibitem{triangles}
The  parametrization (\ref{eq:mixingmatrix}) of the mixing matrix 
suggested originally to describe   the  mixing in
the quark sector,  turned  out  to be  very well suited for neutrinos.
since   solar $\nu$  experiments
essentially measure   $\Delta m^2_{12}$ and the mass components
of  $|\nu_e\rangle$, while atmospheric and reactor experiments
measure $\Delta m^2_{23}$ and the flavor components  of
$|\nu_3\rangle$. In the chosen parametrization
this    corresponds to  the  row  and  column  of the mixing matrix
with the simplest form,   since   the phase  does  not  enter in 
 the expression  for the module of any element.
\bibitem{note} This is  of course an
optimistic  view  of the present  situation,
 since it accepts as  valid interpretations
that are still questioned, and a conservative view because
it ignores the LSND results.
\end{thebibliography}
\end{document}